\documentclass[11pt,a4paper]{article}
\usepackage{jheppub,amsmath,  amssymb,slashed,url,bm,textgreek,upgreek}
\usepackage{graphicx}
\usepackage{epstopdf}

\def\ii{i}

\def\be{\begin{equation}}
\def\ee{\end{equation}}

\def\d{{\mathrm d}}
\def\i{{\mathrm i}}

\def\hat{\widehat}

\def\tilde{\widetilde}

\def\h{\widehat}

\def\d{{\mathrm d}}

\def\R{{\mathbb R}}

\def\[{\bigl [}

\def\]{\bigr ]}

\def\N{{\mathcal N}}

\def\Z{{\mathbb Z}}

\def\h{\widehat}

\def\ii{\iota}

\def\tilde{\widetilde}

\def\bar{\overline}

\def\i{{\mathrm i}}

\font\teneurm=eurm10 \font\seveneurm=eurm7  \font\fiveeurm=eurm5
\newfam\eurmfam
\textfont\eurmfam=\teneurm \scriptfont\eurmfam=\seveneurm
\scriptscriptfont\eurmfam=\fiveeurm

\font\teneusm=eusm10 \font\seveneusm=eusm7 \font\fiveeusm=eusm5
\newfam\eusmfam
\textfont\eusmfam=\teneusm \scriptfont\eusmfam=\seveneusm
\scriptscriptfont\eusmfam=\fiveeusm

\font\tencmmib=cmmib10 \skewchar\tencmmib='177
\font\sevencmmib=cmmib7 \skewchar\sevencmmib='177
\font\fivecmmib=cmmib5 \skewchar\fivecmmib='177
\newfam\cmmibfam
\textfont\cmmibfam=\tencmmib \scriptfont\cmmibfam=\sevencmmib
\scriptscriptfont\cmmibfam=\fivecmmib

\title{Scale And Conformal Invariance in 2d $\sigma$-Models, with an Application to $\N=4$ Supersymmetry}

\author{Georgios Papadopoulos$^1$ and Edward Witten$^2$}
\affiliation{$^1$Department of Mathematics, King's College London, Strand, London WC2R 2LS, UK
\\
$^2$School of Natural Sciences, Institute for Advanced Study,  Princeton, NJ 08540 USA}

\abstract{By adapting previously known arguments concerning Ricci flow and the $c$-theorem, we give a direct proof that in a two-dimensional sigma-model with compact target space,
scale invariance implies conformal invariance in  perturbation theory.   This argument, which applies to a general sigma-model constructed with a target space metric
and $B$-field, is in accord with a more general proof in the literature that applies to arbitrary two-dimensional quantum field theories.
Models with extended supersymmetry and a $B$-field are known to provide interesting test cases for the relation
between scale invariance and conformal invariance in sigma-model perturbation theory.   We give examples showing that in such models,
 the obstructions to conformal invariance suggested by
general arguments can actually occur in models with target spaces that are not compact or complete.   Thus compactness of the target space, or at least a suitable condition of
completeness, is necessary as well as sufficient to ensure that scale invariance implies conformal invariance in models of this type. }

\begin{document}\maketitle

\section{Introduction}\label{intro}

The first part of this article is a general comment on the relation between scale invariance and conformal invariance for two-dimensional sigma-models.
The second part is a specific application to models with extended supersymmetry.

Consider a two-dimensional nonlinear sigma-model on $\R^2$ with target space $M$, described by local coordinates $X^K$ and classical metric $G_{IJ}(X^K)$.   For the case that the only field considered on $M$ is the metric,
the action is
\be\label{sigmaactionintro} I=\frac{1}{4\pi\alpha'} \int_{\R^2} \d^2 x \, G_{IJ}\partial_\alpha X^I \partial ^\alpha X^J\,.\ee
Sigma-model perturbation theory is an expansion in powers of $\alpha'$.   At one-loop order, renormalization requires a counterterm proportional to the Ricci tensor
$R_{IJ}$ of $M$, as originally found by Friedan \cite{friedan}.   However, as also explained by Friedan, one is interested in the metric of $M$ only up to a diffeomorphism of $M$,
and therefore the condition to have a fixed point of the renormalization group is not that $R_{IJ}$ should vanish but rather that a counterterm $\int \d^2x R_{IJ}\partial_\alpha X^I \partial^\alpha X^J$ can be eliminated by a change of coordinates on $M$.   The condition that this counterterm can be eliminated by an infinitesimal coordinate transformation $\delta X^I\sim  V^I(X)$ is
\be\label{fixedpoint} R_{IJ}=D_I V_J+D_J V_I\,. \ee

The condition for conformal invariance of the sigma-model is, however, stronger.   At one-loop order, conformal invariance requires
\be\label{confinv}R_{IJ}=-2D_I D_J \Phi\,,\ee
where $\Phi$ is a scalar field on $M$, the dilaton \cite{CFMP,Ts,Ts2}.   In other words, $V$ must be a gradient:
\be\label{gradv} V_I=- \partial_I \Phi\,,\ee
for some function $\Phi$.    For a detailed explanation of the relation between the conditions (\ref{fixedpoint}) and (\ref{confinv}), as
well as the generalization of these conditions to include a $B$-field\footnote{The present article is concerned only with sigma-model perturbation theory, in which flat $B$-fields (which locally
can be gauged away) are irrelevant.  Hence $B$-fields are always assumed to be non-flat; in other words, $H=\d B$ is always assumed to be non-zero.}
in the sigma-model, see for example \cite{CT}.

As explained by Hull and Townsend \cite{HullTownsend} and Polchinski \cite{Polchinski}, eqn. (\ref{fixedpoint}) is the condition for global scale invariance of the nonlinear sigma-model, while eqn. (\ref{confinv}) is the condition
for local conformal invariance.  This raises the question of whether, in general, scale invariance implies conformal invariance for unitary quantum field theories.   In two dimensions,
Polchinski showed that this question is closely related to the Zamolodchikov $c$-theorem
 \cite{Zam}, which asserts that the central charge monotonically decreases under renormalization group
flow to the infrared.    Indeed, by adapting the original proof of the $c$-theorem, Polchinski was able to give an abstract general proof that
 in a two-dimensional unitary\footnote{Unitarity is a necessary condition as, otherwise, there are counterexamples pointed out by Polchinski in \cite{Polchinski2}, p. 260, and by Itsios, Sfetsos and Siampos in the context of sigma models \cite{Itsiossfetsossiampos}.}
  quantum field theory
with a discrete spectrum of operator dimensions, scale invariance always implies conformal invariance.   This argument did not have a straightforward generalization to dimensions
$d>2$, but eventually it was found  that also in four dimensions the question of relating scale invariance and conformal invariance is  closely related to monotonicity of renormalization group
flow  \cite{DKST}.

In the context of sigma-models, the assumption of a discrete spectrum of operator dimensions holds if the target space $M$ is compact (and smooth, as we will always assume).
One would hope that in sigma-model perturbation theory, one could demonstrate the relation between scale invariance and conformal invariance in a concrete way, rather than relying on
Polchinski's abstract general argument.
For sigma-models
with  compact target space and no $B$-field, Polchinski was able to give such a concrete argument  at one-loop order by
citing an earlier mathematical result of Bourguignon \cite{Bourg}.   Bourguignon had proved that if a metric $G$ on a compact manifold $M$ satisfies $R_{IJ}=D_I V_J+D_J V_I$ for
some $V$, then in fact $R_{IJ}=0$.   Hence the condition for conformal invariance is satisfied, with constant $\Phi$.

This was a nice result,  but it had two drawbacks: it
does not generalize in any obvious way in the presence of a $B$-field, or in higher orders of sigma-model perturbation theory.
Here, we will explain an alternative approach that does give a general result relating scale invariance and conformal invariance for sigma-models, to all orders
and in the presence of a $B$-field.   The argument will be a simple adaptation of
 results of Perelman \cite{Perelman} on the Ricci flow \cite{Hamilton}, as extended  by Oliynyk, Suneeta, and Woolgar \cite{OSW2} to include a $B$-field, in a paper that
 also made the subject much more accessible.  See also \cite{Huhu}.

As in Polchinski's work, we again find that the question of showing that scale invariance implies conformal invariance is closely related to the $c$-theorem.
Indeed,   Perelman's construction gives a way to prove the  $c$-theorem  in sigma-model perturbation theory \cite{OSW,OSW2,Tseytlin2}.  In some ways, this construction
refines the idea of interpreting the central charge of a sigma model as the spacetime effective action \cite{Tseytlin3,CP}.

Perelman was not studying  two-dimensional field theories.  Rather, following a program initiated earlier by Hamilton \cite{Hamilton},
Perelman was using Ricci flow to prove the Poincar\'e conjecture (which asserts that a closed and simply-connected three-manifold
is a three-sphere).
  Ricci flow is the flow on the space of metrics on a manifold $M$ defined by\footnote{\label{alsoimportant} It is also important to slightly modify the flow by
 adding a ``cosmological constant'' term on the right hand side, so that the flow becomes $\frac{\d G_{IJ}}{\d t}=-2 (R_{IJ}-\lambda G_{IJ})$, for some
constant $\lambda$. By giving $\lambda$ a suitable dependence on $t$, one can ensure that the volume of $M$ remains constant under this flow.}
\be\label{zerox} \frac{\d G_{IJ}}{\d t}=-2 R_{IJ}\,, \ee
where $t$ is an auxiliary ``time'' coordinate.

Up to the normalization of the parameter $t$, Ricci flow is the one-loop approximation to the renormalization group flow of the sigma-model;
increasing $t$ represents the flow towards the infrared.   To shorten a rather long story (for a detailed account, see \cite{MorganTian}), to use Ricci flow to prove the Poincar\'e conjecture involves showing that the
Ricci flow tends to simplify the geometry.  This could not happen if the Ricci flow has periodic orbits.  So a step toward understanding Ricci flow and eventually
proving the Poincar\'{e} conjecture was to prove that  Ricci flow is in some sense monotonic, with no periodic orbits.   In proving this, Perelman  arrived at what can be
 interpreted \cite{OSW,OSW2,Tseytlin2} as a proof of the $c$-theorem in sigma-model perturbation theory.

Perelman also had to consider metrics that satisfy $R_{IJ}=D_I V_J+D_J V_I $ for some vector field $V$ on $M$.
Mathematically,  manifolds endowed with such metrics -- or for brevity sometimes the metrics themselves -- have been called (steady\footnote{Metrics that satisfy the more general condition $R_{IJ}-\lambda G_{IJ}=D_I V_J+D_J V_I$, for some constant $\lambda$, are also important. They are called expanding or shrinking Ricci
solitons depending on the sign of $\lambda$.})
Ricci solitons; they were first studied in \cite{Hamilton2}.   If $V$ is the gradient of a scalar function, the Ricci soliton is called a gradient Ricci soliton.
Ricci solitons  are important because they are associated to singularities that can develop under Ricci flow;  for example, see the exposition in \cite{Bamler}.
Such singularities  present the main technical difficulty in using Ricci flow to prove the Poincar\'e conjecture, so as steps toward that conjecture,
Perelman proved a number of results on Ricci solitons.

In general, Perelman had to consider Ricci solitons that are complete but not necessarily compact.   However, for our purposes, Perelman's relevant result
was a new proof that  any compact  (steady) Ricci soliton is a gradient, or in other words, in terms of sigma-models with compact target space and no $B$-field, scale invariance implies conformal invariance.
For a useful short explanation of Perelman's argument, see the proof of Proposition 2.1 in \cite{Cao}.

 Perelman's proof that compact Ricci solitons are gradients extends naturally to the case that a $B$-field is incorporated in the sigma-model.
 That is not  surprising, since it has been known that the $B$-field can be included in a Perelman-style proof of the $c$-theorem \cite{OSW2}.
Moreover, the argument based on Perelman's approach extends to all orders of
sigma-model perturbation theory.

Thus, a Perelman-style argument gives a satisfactory understanding of the relation between scale invariance and conformal invariance in the context of sigma-model
perturbation theory.   Our goal in section \ref{Perelmanstyle} of this article, accordingly, will be to explain Perelman's argument and its generalization to include the $B$-field, along the
lines of  \cite{OSW2}.  After reviewing the Perelman-style proof of the $c$-theorem, we will explain the (short) additional
step that is needed to establish the relation between scale invariance and conformal invariance.

In section \ref{application}, we will  reconsider a specific class of examples of the relation between scale invariance and conformal invariance in sigma-model perturbation theory.   This
involves two-dimensional sigma-models with $\N=4$ supersymmetry and a non-trivial  $B$-field.
 For early work on such models, see
 \cite{Curtrightzachos}-\cite{Hullreview}.  The last citation also contains a review of much of the early literature.
For sigma-models with vanishing $B$-field, $\N=4$ worldsheet supersymmetry
is associated to hyper-K\"{a}hler geometry and conformal invariance.
 In the presence of a $B$-field, matters are not so simple.
The one-loop beta functions of a general sigma-model with extended supersymmetry and a $B$-field were originally analyzed in \cite{Buscher, Hull} and further analyzed in
\cite{HullTownsend}.
It turns out that $(0,4)$ supersymmetry of such
a $\sigma$-model (or $(0,2)$ supersymmetry with a generalized Calabi-Yau condition)   implies
one-loop finiteness of the theory formulated on $\R^2$. One-loop finiteness is the statement that the one-loop counterterms
(which are proportional to the $\beta$ functions of the metric and the $B$-field but not of the dilaton) vanish up to a field redefinition.
But, as was soon understood \cite{HullTownsend}, in the presence of a $B$-field, one-loop finiteness is not strong enough to imply
conformal invariance; rather, it implies a condition that generalizes the Ricci soliton equation (\ref{fixedpoint}) to include a $B$-field.
To put it differently, in perturbation theory, $(0,4)$ supersymmetry (or $(0,2)$ supersymmetry and a Calabi-Yau condition)
with a $B$-field leads naturally to scale invariance without conformal invariance.

The question has a few variants, depending on exactly what properties one assumes for $M$.   One may impose conditions on $M$ that lead to extended
supersymmetry for only right-movers (or left-movers) of the sigma-model, or one may impose conditions that lead to extended supersymmetry for both left- and right-moving
modes.
 The extension may be to either $\N=2$ or $\N=4$ supersymmetry.   And, in the case of extended supersymmetry for both left- and right-moving modes,
the complex structures associated to left- and right-moving extended supersymmetry may commute or not commute, a distinction that will be important in this article
and whose importance was observed in
\cite{JGCHMR}.  We describe some of the options in sections \ref{KT}-\ref{worldsuper}.  Much of this material is a review, but some statements about the Lee form for
particular types of geometry are new.  The Lee form is important because it controls the potential obstruction to conformal invariance in scale-invariant models
with extended supersymmetry.   In section \ref{examples}, we describe some simple and relatively well-known examples,  and in
 section \ref{scalerevisited}, we reconsider the relation between scale invariance and conformal invariance
in light of these examples.
Once a model satisfies the one-loop condition for scale invariance, the one-loop condition for conformal invariance is satisfied if and only if there exists a dilaton field $\Phi$
on the sigma-model target space $M$ that satisfies a certain condition.   The basic conclusion in section \ref{scalerevisited}
is that the ability to satisfy this condition is no better than is predicted by the constraints discussed in  sections \ref{KT}-\ref{worldsuper}.
For example, in a generic model with a $B$-field and $(4,4)$ (or less) supersymmetry, there is a local obstruction to finding a dilaton field $\Phi$ on $M$ with the properties
that lead to conformal invariance.   If we suppose that the complex structures related to supersymmetry of left- and right-movers commute, then there is no local obstruction
to  conformal invariance, but there can still be a global obstruction.    Notably, and in keeping with the general argument in \cite{Polchinski} as well as the specific
sigma-model argument that we explain in section \ref{Perelmanstyle}, the examples that we can find that realize the obstruction have target spaces that are singular and
not complete.  A condition of compactness or at least some sort of completeness is necessary as well as sufficient to ensure that in these models,
scale invariance implies conformal invariance.

Another perspective on the geometries that we will discuss in section \ref{application} is provided by generalized complex geometry \cite{Hitchin}.
For example, the bihermitian -- or in the language we use in section \ref{application}, bi-KT -- geometries that were introduced in \cite{JGCHMR} have
been reinterpreted in terms of generalized K\"{a}hler geometry \cite{Gualtieri}.  This approach might very well lead to further results about the Lee form,  but will not be considered in the present article.    When the article was substantially
finished, we learned of recent work \cite{GFS}  in which the Ricci flow -- generalized to include the $B$-field as in \cite{OSW2} -- is studied from the point of view of generalized geometry.   Our result on the relation between conformal invariance and scale invariance in leading order is actually equivalent to Corollary 6.11 in that reference.

\section{Scale and Conformal Invariance in Sigma-Model Perturbation Theory}\label{Perelmanstyle}

\subsection{Preliminaries}\label{preliminaries}

The fields that are common to all closed-string theories are the metric $G_{IJ}$, with Riemann tensor $R_{IJKL}$, the $B$-field, which is a two-form with three-form field strength
$H=\d B$,
and the dilaton $\Phi$. (It is straightforward to include in the following also the massless gauge fields of the heterotic string, or a ``cosmological constant'' term proportional
to $D-26$, where $D$ is the target space dimension.   Either of these additions just modifies the potential in the effective Schr\"{o}dinger equation that we will encounter.)
In Euclidean signature, up to an overall constant that can be absorbed in an additive shift of $\Phi$, the standard action for these fields, in leading order
in $\alpha'$, is
\be\label{leadact}S(G,B,\Phi)=\int \d^D X \sqrt G e^{-2\Phi}\left(-R -4 G^{IJ}D_I\Phi D_J\Phi  +\frac{1}{12} H^2\right)\,. \ee
A noteworthy fact about this action is that the kinetic energy $G^{IJ}D_I\Phi D_J\Phi $ of the dilaton field has a coefficient of the wrong sign.   That is not true of the other fields.   The $H^2$ term in the action
has a positive coefficient, and in Euclidean signature the standard Einstein-Hilbert action is a negative multiple of the scalar curvature $R$.

Before concluding that the theory must be pathological, we should pause and note that pure Einstein gravity has the same feature \cite{GHP}.  Indeed, if one expands the
Einstein-Hilbert action $-\int\d^D X \sqrt G R$ around a classical solution, one finds that the kinetic energy for the conformal mode of the metric, that is for a rescaling of the metric
by $G\to e^\phi G$, is negative.

Going back to the string theory case, if the action $S$ is expanded around a classical solution, one finds that just one linear combination of the two modes $\phi$ and $\Phi$ has
a kinetic energy of the wrong sign.    The kinetic energy for the two modes $\phi$ and $\Phi$ is associated to a quadratic form with one positive and one negative
eigenvalue.

In Einstein gravity, Gibbons, Hawking, and Perry  \cite{GHP} suggested to deal with the wrong sign kinetic energy of the conformal mode by a Wick rotation $\phi\to \i\phi$.   This amounts roughly to saying
that one should maximize the action as a function of $\phi$ instead of minimizing it.

Carrying over this idea to the string theory case, one might arrive at Perelman's idea of, roughly, maximizing $S$ as a function of $\Phi$, to obtain a well-behaved functional of the
other fields.   To see that this could work, consider a quadratic form $Q(x,y)= ax^2+2bxy+cy^2$ in two variables $x,y$, with coefficients $a,b,c$ chosen so that $ac-b^2<0$,
ensuring that $Q$ has indefinite signature.   If $c<0$, we cannot minimize $Q$ as a function of $y$ for fixed $x$, but we can maximize it.   The maximum of $Q$ for fixed $x$
is $\frac{1}{c} (ac-b^2)x^2$.   Since $c$ and $ac-b^2$ are both assumed negative, this is a positive function of $x$.

Before discussing in what sense one can maximize $S$ as a function of $\Phi$, we follow Perelman \cite{Perelman} as interpreted and generalized to
include the $B$-field in \cite{OSW2}, and rewrite $S$ as follows:
\be\label{newact}S(G,B,\Phi)=-\int\d^DX \sqrt G e^{-\Phi}\left(-4 \Delta^2  +R-\frac{1}{12}H^2\right) e^{-\Phi}\,.\ee
Here $\Delta^2=G^{IJ}D_I D_J$ is the scalar Laplacian.   We do not literally want to maximize $S$ as a function of $\Phi$, because the maximum would be attained for
$\Phi\to \pm \infty$ and would be 0 or $ \infty$, depending on the other fields.  Instead Perelman's approach is to maximize $S$ as a function of $\Phi$ under the constraint\footnote{We consider compact
manifolds only, so that $\int \d^D X\sqrt G e^{-2\Phi}<\infty$.   However, see \cite{OSW} for a variant of this argument suitable for complete Riemannian manifolds of infinite volume.}
 $\int \d^DX \sqrt G e^{-2\Phi}=1$.
Including a Lagrange
multiplier $\lambda$, a maximum of $S$ under the constraint is an extremum of
\be\label{extact}S' = -\int\d^DX \sqrt G e^{-\Phi}\left(-4 \Delta^2  +R-\frac{1}{12}H^2\right) e^{-\Phi} +\lambda\left( \int \d^D X \sqrt G e^{-2\Phi} -1\right)\,. \ee
Extremizing this with respect to $\Phi$, we learn that $e^{-\Phi}$ must be an eigenfunction of the Schr\"{o}dinger-like operator $-4 \Delta^2  +R-\frac{1}{12}H^2$ with
eigenvalue $\lambda$:
\be\label{eival}\left(-4 \Delta^2  +R-\frac{1}{12}H^2\right)e^{-\Phi}=\lambda e^{-\Phi}\,.\ee
Of course, $e^{-\Phi}$ is positive-definite.   A Schr\"{o}dinger-like operator has a unique normalized positive-definite eigenfunction, namely the ground state.   This ground state is the unique
extremum  of $S$ under the constraint, and in particular it is the absolute maximum.

Thus  the constrained maximum of $S(G,B,\Phi)$ as a function of $\Phi$ for fixed $G$ and $B$ always exists and is unique.  Because of this uniqueness, and general properties of
elliptic partial differential equations, the maximum varies smoothly as a function of $G$ and $B$.
Therefore,   $\bar S(G,B)=S(G,B,\Phi(G,B))$, where
$\Phi$ is  regarded as a function $\Phi(G,B)$,
evaluated at the constrained maximum of $S$, is a smooth function of $G$ and $B$.

\subsection{Monotonicity of Renormalization Group Flow}\label{monotone}

To see that $\bar S(G,B)$ has the properties of a $c$-function is a simple consequence of standard formulas for the relation between the variation of $S(G,B,\Phi)$ and the
beta functions.   There is one important detail, however.   It is most useful to begin with the formula for the variation of $S$ under a change of $G$ and $B$, keeping fixed
not $\Phi$ but $\h\Phi=\Phi-\log \sqrt[^4]G$.  Though we do not have a completely satisfactory {\it a priori} explanation for why this procedure gives the best results,
the following is suggestive.

In Perelman's approach, $e^{-\Phi}$ is treated as a quantum mechanical wavefunction, a vector in a Hilbert space of square integrable functions on the spacetime $M$.
However, there is not really a natural Hilbert space of $L^2$ {\it functions}; the inner product in this Hilbert space depends on the measure $\d^DX\sqrt G$, which depends
on the metric $G$.   Instead, on any smooth manifold $M$ there is a completely natural Hilbert space of $L^2$ {\it half-densities}.   A half-density is the square root of a measure or density;
in other words, it transforms under coordinate changes as the square root of a density.   In the present context, $e^{-\Phi}$ is a function but $e^{-\h\Phi}=e^{-\Phi}\sqrt[^4]G$
is a half-density.   The norm of the half-density $e^{-\h\Phi}$ is simply $\int_M \d^D X e^{-2\h\Phi}$ with no structure required to make sense of this integral beyond the fact that
$M$ is a smooth manifold.\footnote{In fact, a smooth structure on $M$ is more than is needed here.}   It turns out that the most useful procedure is to vary $G$ and $B$ keeping
fixed not the function $\Phi$ but the quantum state $e^{-\h\Phi}$.

Bearing in mind that we want to vary $G$ and $B$ keeping fixed $\h\Phi$ rather than $\Phi$, the standard relation
 between the action $S$ and the sigma-model $\beta$ functions can be written:
\begin{align}\label{betafns}\left. \frac{\delta S}{\delta G_{IJ}}\right|_{\h\Phi}&
= e^{-2\h\Phi} \beta_G^{IJ} = e^{-2\h\Phi} \left(R^{IJ}-\frac{1}{4} H^I{}_{KL}H^{JKL} +2 D^I D^J\Phi\right)\cr
                       \left.    \frac{\delta S}{\delta B_{IJ}}\right|_{\h\Phi}& =e^{-2\h\Phi} \beta_B^{IJ}= e^{-2\h\Phi} \left(\frac{1}{2} D_K H^{KIJ} -D^K \Phi H_{KIJ}\right)\,. \end{align}
                           Here $\beta_G$ and $\beta_B$ are the beta functions that appear in the renormalization group equation for the evolution of $G$ and $B$:
   \begin{align}\label{zembo}
   \frac{\d G_{IJ}(X)}{\d t}&=-\beta_{G, IJ} (X) \cr
                                              \frac{\d B_{IJ}}{\d t}& =-\beta_{B,IJ}(X)\,.
   \end{align}
There is a similar formula for $\delta S/\delta\Phi$, but we do not need it as we will be evaluating $S$ at an extremum with respect to $\Phi$.
Note that extremizing $S$ with respect to $\Phi$ at fixed $G$ and $B$ is the same as extremizing $S$ with respect to $\h\Phi$ at fixed $G$ and $B$.
From eqn. (\ref{betafns}), it follows that
\begin{align}\label{otherfns} \frac{\delta \bar S}{\delta G_{IJ}}&=  e^{-2\h\Phi} \beta_G^{IJ} \cr
\frac{\delta \bar S}{\delta B_{IJ}} & = e^{-2\h\Phi} \beta_B^{IJ}\,.     \end{align}
This is true because $\bar S(G,B)=S(G,B,\Phi(G,B)),$ where $\Phi(G,B)$ is such that
\be\label{critpt}\left.\frac{\delta S(G,B,\Phi)}{\delta \Phi} \right|_{\Phi=\Phi(G,B)}=0\,.\ee
A statement equivalent to eqn. (\ref{otherfns}) is that in a first order variation of $G$ and $B$, one has
\be\label{firstvar}\delta \bar S=\int \d^DX e^{-2\h\Phi}\left( \beta_G^{IJ}\delta G_{IJ} +\beta_B^{IJ}\delta B_{IJ}\right)\,. \ee

The monotonicity of renormalization group flow is an immediate consequence of eqns.     (\ref{zembo}) and (\ref{firstvar}):
\be\label{embo}\frac{\d \bar S}{\d t} = -\int \d^DX  e^{-2\h\Phi}\left( \beta_{G,IJ}\beta^{G,IJ}+\beta_{B,IJ}\beta^{B,IJ}\right)\leq 0\,,    \ee
This establishes  a version of the $c$-theorem, to lowest order in sigma-model perturbation theory.

\subsection{Why Not ``Einstein'' Flow?}\label{einstein}

Before going on, let us ask why, from Perelman's point of view, it was necessary to include the dilaton in this analysis.      Perelman wanted a monotonicity result for the Ricci flow
\be\label{riccif}\frac{\d G_{IJ}}{\d t}=- 2R_{IJ}\,. \ee
It would have been ideal if this could be interpreted as gradient flow with respect to a multiple of the Einstein-Hilbert action $-\int \d^D x \sqrt G R$.   However, that is not the case;
gradient flow with respect to the Einstein-Hilbert action would give instead
\be\label{niccif} \frac{d G_{IJ}}{\d t}=-2 \left(R_{IJ}-\frac{1}{2} G_{IJ} R\right)\,. \ee
But why did Hamilton \cite{Hamilton} and following him Perelman \cite{Perelman} not consider the ``Einstein'' flow (\ref{niccif}) instead of the Ricci flow (\ref{riccif})?   The answer
to this question is precisely that the conformal mode appears in the Einstein-Hilbert action with the wrong sign.   The Ricci flow equation (\ref{riccif}) is a parabolic differential equation,
similar to the heat equation.   Integrated forward in time, it
 tends to smooth irregularities in the metric, evolving in the direction of simplification (once certain singularities
that present the main difficulties in the theory are excised).      Because the conformal mode enters the Einstein-Hilbert action and the corresponding equations of motion with the
wrong sign, the ``Einstein'' flow equation has the opposite effect on the conformal mode: it magnifies irregularities.  Starting with generic smooth initial data, the ``Einstein'' flow equation
cannot be integrated forward in time for any non-zero time interval, just as the heat equation and the Ricci flow equation cannot be integrated backwards in time.

So one would want to to eliminate the conformal mode before interpreting the ``Einstein'' flow as a gradient flow.   Could one do this by maximizing the Einstein-Hilbert action with respect
to the conformal mode, under a constraint on the volume, similarly to what Perelman did with the dilaton in the case of the string effective action?   Trying to do this leads to the
Yamabe problem, a celebrated and much-studied problem in differential geometry \cite{LeeParker}.   There is a very elegant general theory of the Yamabe problem, but for the purposes of
finding a $c$-theorem, this theory is not very satisfactory: in general, the volume-constrained
extremum of the Einstein-Hilbert action with respect to the conformal factor is not unique, and the absolute
maximum can jump discontinuously as the metric is varied.   The crucial step in Perelman's work that would not go through in the absence of the dilaton is that the problem of constrained
maximization of the action with respect to the wrong-sign mode always has a unique and smoothly varying solution.

\subsection{Scale Invariance and Conformal Invariance}\label{scaleconf}

Finally we will explain the small additional step that is needed to explain the relation between scale-invariance and conformal invariance.
The condition for scale-invariance is not that $\beta_G$ and $\beta_B$ must vanish, but that they must vanish up to an infinitesimal diffeomorphism generated by a vector field $V$ along
with a $B$-field gauge transformation generated by a one-form $\Lambda$.
Under the action of $V$ and $\Lambda$, the variation of $G$ and $B$ is
\begin{align}\label{timely}     \delta G_{IJ}=D_I V_J+D_JV_I\,,~~~~~\delta B_{IJ}=V^K H_{IJK}+\partial_I\Lambda_J-\partial_J \Lambda_I\,.                           \end{align}
Thus the condition for global scale-invariance is
\be\label{scalecond}\beta_{G,\,IJ}=D_I V_J+D_JV_I\,,~~~~~~~\beta_{B\, IJ}=V^K H_{IJK} +\partial_I\Lambda_J-\partial_J\Lambda_I\,.\ee
The condition for local conformal invariance is instead $\beta_G=\beta_B=0$, in other words, the same condition but with $V=\Lambda=0$.  By the Curci-Paffuti relation \cite{CP}, if
$\beta_G=\beta_B=0$, then $\beta_\Phi$ is a constant (the central charge) and conformal invariance holds.  So to demonstrate that scale invariance implies conformal
invariance in the lowest order of sigma-model perturbation theory, we have to show that if the condition (\ref{timely}) for global scale invariance holds, then in fact
the same condition holds with $V=\Lambda=0$ and the model is conformally invariant in this order.

This is an immediate consequence of facts already stated plus the fact $\bar S$  is invariant under diffeomorphisms and $B$-field gauge transformations.
   That invariance of $\bar S$
means that $\delta \bar S$ vanishes in first order  if $G$ and $B$ are varied as in eqn. (\ref{timely}).   Since in general $\delta\bar S$ is given by eqn. (\ref{firstvar}),
the condition that it vanishes if  the variations $\delta G$ and $\delta B$ are those produced by a vector field $V$ and a one-form $\Lambda$ gives
\be\label{diffvar} 0 =\int_M \d^DX  e^{-2\h\Phi} \left( \beta_G^{IJ}(D_I V_J+D_JV_I) +\beta_B^{IJ} \left(V^K H_{IJK}+\partial_I\Lambda_J-\partial_J \Lambda_I \right)\right)\,. \ee
This is just a statement of diffeomorphism and gauge invariance   and holds irrespective of any equations of motion.   In a globally scale-invariant model, we have eqn. (\ref{scalecond}), and
therefore
\be\label{iffvar} 0=\int_M \d^D X e^{-2\h\Phi} \left(\beta_G^{IJ}\beta_{G\,IJ} +\beta_B^{IJ}\beta_{B\,IJ}\right)\,.\ee
This implies that $\beta_G=\beta_B=0$, and conformal invariance holds.

\subsection{Higher Orders}\label{higherorders}

Both for the $c$-theorem and for the relation between scale invariance and conformal invariance, we have presented up until this point arguments that are valid in lowest
order of sigma-model perturbation theory.   However, the nature of the construction is such that the conclusions automatically apply to all orders.

We used two facts: the relation between the spacetime effective action $S(G,B,\Phi)$ and the sigma-model beta functions, and the existence of a unique maximum
of $S(G,B,\Phi)$ as a function of $\Phi$ keeping $G$ and $B$ fixed.  The relationship between the effective action and the beta functions holds to all orders \cite{CP}.  (See \cite{AA}
for a recent discussion.)   As for the existence and uniqueness of a maximum of the spacetime effective
action as a function of $\Phi$, because the maximum exists and is unique and nondegenerate
in lowest order of perturbation theory, it  automatically  persists (with perturbative modifications) to all orders  in perturbation theory.

To elaborate slightly on this point, in general consider a set of real variables $\vec\phi=(\phi_1,\cdots, \phi_k)$ and a system of possibly nonlinear
equations $F_i(\vec\phi)=0$, $i=1,\cdots,k$.
A solution $\vec\phi=\vec\phi_0$ of the equations $F_i(\vec\phi)$ is said to be nondegenerate if $\left.\det\frac{\partial F_i}{\partial \phi_j}\right|_{\vec\phi=\vec\phi_0}\not=0$.
Equivalently, a solution is nondegenerate if the linear equation that one gets by linearizing the equation $F(\vec\phi)=0$ around the given solution has no zero-mode.
In general, if a solution $\vec\phi=\vec\phi_0$ of an equation $F(\vec\phi)=0$ is nondegenerate, then the existence of this solution is stable against sufficiently small
perturbations to $F$: as $F$ is varied, the solution $\vec\phi=\vec\phi_0$ will move but, for sufficiently small perturbations and in particular in perturbation theory, will not disappear.

If instead of a finite set of variables, we consider a field $\Phi(X)$ on some manifold $M$,
the same statement holds for a solution of an elliptic differential equation $F(\Phi(X))=0$.   A solution is called nondegenerate if the linear equation obtained by linearizing around
the given solution has no nonzero solution.   A nondegenerate solution is always stable in the sense that it varies smoothly and remains nondegenerate under a suitable class
of variations of $F$.  In particular, this is true for arbitrary perturbative modifications of $F$.

In the preceding analysis, the equation that is  solved to maximize the effective action as a function of $\Phi$ was the equation that says that $e^{-\Phi}$ is the normalized
positive ground state wavefunction of a certain Schr\"{o}dinger operator.   The Schr\"{o}dinger equation is linear, and nondegeneracy   of the solution for $\Phi$
is just the well-known fact that the ground state of a Schr\"{o}dinger operator
of the sort that we encountered is nondegenerate.

Therefore, the conclusions about the $c$-theorem and the relation between scale invariance and conformal   invariance hold to all orders of sigma-model perturbation theory.
The reason that we do not make a claim beyond perturbation theory is primarily that although a sigma-model with target $M$ can be constructed order by order in perturbation theory
for any $M$, the conditions under which this sigma-model exists nonperturbatively are  in general not fully understood.   Hence it appears that one does not have a satisfactory
framework to make a nonperturbative claim in general.

\section{Geometry and Worldsheet Supersymmetry}\label{application}

 Worldvolume supersymmetry in a sigma-model with target space $M$  is intimately related to  the geometry of $M$.  For a generic Riemannian target space $M$, one can construct
 a model with $(1,1)$ supersymmetry.  In the absence of a $B$-field, this is promoted to $(2,2)$ supersymmetry if $M$ is K\"{a}hler \cite{Zumino} and to $(4,4)$ supersymmetry if $M$
 is hyper-K\"{a}hler \cite{LAGDF}. See also \cite{Curtrightfreedman}.  Under the same conditions, $(0,1)$ supersymmetry is  promoted to $(0,2)$ or $(0,4)$.

   In the presence of a $B$-field, there is a richer and more complicated story, leading to the construction of supersymmetric sigma-models with
 skew-symmetric torsion $H=\d B$ and with $(p,q)$ supersymmetry for various values of $(p,q)$.   For early work on such models, see  \cite{Curtrightzachos}-\cite{Hullreview}.

   For our purposes in this article, what is relevant is that, as observed long ago \cite{HullTownsend,Polchinski},
supersymmetric sigma-models with torsion
provide an interesting test case for the relation between scale invariance and conformal invariance.      In sections \ref{KT}-\ref{examples}, we explain some  background on
 such models and  describe some simple examples.   In section \ref{scalerevisited}, we then consider the relationship between scale invariance
 and conformal invariance in the context of these examples.    The basic conclusion is as predicted  in \cite{Polchinski} and is in accord with the  explicit proof that we have given in section
 \ref{scaleconf} for sigma-models with a $B$-field: although geometries can be constructed that satisfy
 the perturbative conditions for scale invariance without conformal invariance, the examples that we are able to construct are non-compact and singular.

\subsection{K\"ahler Geometry with Torsion}\label{KT}

A hermitian manifold is a Riemannian manifold $M$  of even dimension $D=2n$ with a metric $G$ and an integrable complex structure $I$, such that $G$ is hermitian  with respect to $I$.
A complex structure $I$ is in particular a tensor $I^P{}_Q$ satisfying $I^P{}_Q  I^Q{}_R=-\delta^P{}_R$, or equivalently $I^2=-1$.   Integrability means that locally $M$ admits
complex coordinates $z^1,\cdots, z^n$  that are holomorphic with respect to $I$;  transition functions relating such descriptions are also  holomorphic with respect to $I$.
In such coordinates, $I$ is a constant tensor:
 \be
 I=I^P{}_Q \partial_P\otimes dx^Q= \i \delta^\alpha{}_\beta \partial_\alpha\otimes dz^\beta-\i \delta^{\bar\alpha}{}_{\bar\beta} \partial_{\bar\alpha}\otimes dz^{\bar\beta}\,.
 \ee
The condition for the metric to be hermitian is $G_{PQ} I^P{}_K I^Q{}_L=G_{KL}$; equivalently, $G$ is of type $(1,1)$, meaning that its nonzero elements in the basis
$\d z^\alpha$, $\d\bar z^{\bar \beta}$ are of the form $G_{\alpha\bar\beta}=G_{\bar\beta\alpha}$.  Associated with $I$ and $G$ is the two-form $I=\frac{1}{2}I_{PQ}\, \d x^P\wedge\d x^Q$,
where $I_{PR}=G_{PQ} I^Q{}_R$.   In local complex coordinates,  $I_{\alpha\bar\beta}=-\i G_{\alpha\bar\beta}$, $I_{\bar\beta \alpha}=\i G_{\bar\beta\alpha}$; thus in particular, $I$
is of type $(1,1)$.
Note that we use the same symbol $I$ for the complex structure and the associated two-form; hopefully the context will always make
clear what is intended.   On the complex manifold $M$, the exterior derivative $\d$ has the usual decomposition $\d=\partial+\bar\partial$, where $\partial$ and $\bar\partial$
are of respective types $(1,0)$ and $(0,1)$.

A connection is said to be compatible with the hermitian data if $G$ and $I$ are covariantly constant.  Such a connection necessarily has holonomy in  $U(n)$.
The Riemannian connection generically does not have this property (as $I$ is not covariantly constant, unless $M$ is K\"{a}hler), but there are several natural connections
for which both $G$ and $I$ are covariantly constant.  Of interest here is the unique such connection for which the torsion is completely antisymmetric and thus is given
by a three-form $H$.   This connection, which we call $\hat\nabla,$ can be explicitly defined in terms of the Riemannian connection $\nabla$ by specifying the covariant
derivative of an arbitrary vector field $V$:
\be\label{condef}\hat \nabla_R V^S=\nabla_R V^S+\frac{1}{2} H^S{}_{RT} V^T, ~~~~H^S{}_{RT}=G^{SS'} H_{S'RT}\,. \ee
Covariant constancy of $G$ and $I$ is the assertion that  $\hat\nabla_K G_{LP}=\hat\nabla_K I^L{}_P=0$, or more briefly $\hat\nabla G=\hat\nabla I=0$. To satisfy $\hat \nabla I=0$,
$H$ must be  determined from $G$ and $I$ as
\begin{eqnarray}
H= - \ii_I \d I=-\d_I I\,,
\label{torsion}
\end{eqnarray}
where $\d I$ is the exterior derivative of the Hermitian form $I$, $\ii_I$ is the inner derivation\footnote{The inner derivation $\ii_I L$ of a $k$-form $L$ with respect to $I$ is $\ii_I L=\frac{1}{(k-1)!} I^K{}_{P_1} L_{K P_2\dots P_k} \d x^{P_1}\wedge\dots\wedge \d x^{P_k}$.} with respect to $I$, and  $\d_I=\ii_I \d-\d \ii_I=\i(\partial-\bar\partial)$. The 3-form $H$ is of type $(2,1)\oplus (1,2)$  with respect to the complex structure $I$, i.e. its  (3,0) and (0,3) components vanish. This is a consequence of $\hat\nabla I=0$ together with the integrability of the complex structure $I$.

In a generic Hermitian geometry $(M; G,I)$,  the three-form $H$ is not closed,  $\d H\not=0$.
However, in the application to sigma-models, at least in the context of Type II superstring theory,\footnote{In the heterotic string, one has $H=\d B+{\mathrm {CS}}$ where ${\mathrm {CS}}$
is a Chern-Simons form.  Hence in general, $\d H\not=0$.  We will not consider that generalization in the present article.   For an early discussion of the relation of scale and
conformal invariance in that context, see \cite{HullHeterotic}.}
$H$ is interpreted as the three-form field strength of a two-form field $B$: $H=\d B$.   For this, $H$ must be closed
(and additionally  must satisfy a Dirac quantization condition).
A manifold $M$ endowed with the pair $G,I$ has been called a KT manifold (K\"{a}hler with torsion), and if $\d H=0$, then  $M$ is called a strong KT manifold.
   The condition $\d H=0$ is equivalent to \begin{eqnarray}
\partial\bar\partial I=0\,.
\end{eqnarray}
Thus, locally $I=\partial \bar Y+\bar\partial Y$, where the  (1,0)-form $Y$  is the closest analog in this situation of the K\"{a}hler potential of a K\"{a}hler manifold \cite{Hullwitten}.
 Of course if $H=0$, then $(M; G,I)$ is a K\"ahler manifold.

The Lee form of a KT manifold is defined as
\begin{equation}
\hat\theta_K\equiv D^P I_{PQ} I^Q{}_K=-\frac{1}{2} I^L{}_K H_{LPQ} I^{PQ}\,,
\label{leeform}
\end{equation}
where the second equality follows as a consequence of $\hat\nabla I=0$. The Lee form $\hat\theta$ depends on the complex structure $I$.

The Riemann tensor of the connection with torsion $\hat\nabla$  is defined  by
\begin{equation}\label{defriemann}
\hat\nabla_P \hat\nabla_Q V^S-\hat\nabla_Q \hat\nabla_P V^S=\hat R_{PQ}{}^S{}_T V^T- H^T{}_{PQ} \hat\nabla_T V^S\,,
\end{equation}
for any vector field $V$.
However, $\h R_{PQ}{}^S{}_T$ does not satisfy all of the algebraic properties of the ordinary Riemann tensor of Riemannian geometry.   If we define
$\h R_{PQST}=G_{SS'} \h R_{PQ}{}^{S'}{}_{T}$, then $\h R_{PQST}$ is obviously antisymmetric in the first two indices, and the fact that $G$ and $I$  are covariantly constant with respect
to $\h\nabla$ implies that $\h R_{PQST}$ is antisymmetric and of type $(1,1)$ in the last two indices.   The type $(1,1)$ condition is equivalent to
\be\label{typeoneone} \h R_{PQST}=\h R_{PQS'T'}\,I^{S'}{}_S I^{T'}{}_T\,. \ee
The Bianchi identities satisfied by $\h R_{PQST}$ are described in
appendix \ref{ap-a}.

A generic K\"{a}hler manifold is, of course, not Ricci-flat and does not satisfy the condition of scale or conformal invariance in leading order of sigma-model perturbation theory.
The K\"{a}hler manifolds that do satisfy this condition are  the Calabi-Yau manifolds.   Similarly, the strong KT manifolds
that we have investigated so far generically do not satisfy the condition for leading order scale or conformal invariance.
A condition that ensures that the leading order condition for scale invariance is satisfied is the analog of the Calabi-Yau condition.\footnote{For a local analysis of the
implications of the Calabi-Yau condition in this context, see \cite{HullLocal}.   Eqns. (6.68) and (6.69) in that paper correspond to eqns. (\ref{meldo}) and (\ref{weldo}) below.}
Restricting the holonomy of  $\hat\nabla$ to be contained in $SU(n)\subset U(n)$, one gets what has been called a strong CYT manifold (Calabi-Yau with torsion)
or a generalized Calabi-Yau manifold.
The Riemann tensor  of a strong CYT manifold obeys an additional condition
\be\label{addcond}
\h R_{PQST}\, I^{ST}=0\,.
\ee
Eqns. (\ref{typeoneone}) and (\ref{addcond})  together with the Bianchi identity (\ref{b1}) and eqn. (\ref{leeform}) yield
\begin{eqnarray}\label{combined}
\hat R_{KL}=\hat\nabla_K \hat\theta_L\,.
\label{riccilee}
\end{eqnarray}
  The symmetric and antisymmetric parts of this equation are actually equivalent to the conditions (\ref{scalecond}) for global
scale-invariance, with $V$ being the vector field dual to $\h\theta$ and $\Lambda=\h\theta$.   This (or more precisely the one-loop finiteness of the sigma-model, which is a closely related statement)
was first recognized in \cite{Buscher} for a particular class of examples and in \cite{Hull} for the general case of
 a strong CYT manifold.
Leading order conformal invariance holds if and only if there is a scalar function $\Phi$ (the dilaton) such that
\be\label{zeldo} \hat\nabla_K\hat\theta_L =-2\hat\nabla_K\partial_L\Phi\,.\ee
This condition follows from but is weaker than $\hat\theta_L=-2\partial_L\Phi$.   It says that if we define
\be\label{meldo} \h V_L=\hat\theta_L+2\partial_L\Phi\,,\ee
then $\h V_L$ is covariantly constant for the connection $\h\nabla$:
\be\label{weldo} \hat \nabla_K \h V_L=0\,.\ee
Concretely, given the definition of $\h\nabla,$ the part of this equation that is symmetric in $K$ and $L$ says that $\h V^L$ is a Killing vector field, satisfying $D_K \h V_L+D_L \h V_K=0$,
and the antisymmetric part
says that $\h V^P H_{PQR}+\partial_Q \h V_R-\partial_R \h V_Q=0$.
Together these relations say that
a diffeomorphism generated by the vector field $\h V^P$ combined with a $B$-field gauge transformation generated by the one-form $\Lambda_L=\h V_L$
leaves invariant the sigma-model background fields $G,B$.  If $\h\nabla_K \h V_L=0$, then in eqn. (\ref{combined}), we can replace $\h\theta_L$ with $-2\partial_L\Phi$, and eqn. (\ref{combined})
becomes the one-loop condition of conformal invariance with dilaton $\Phi$. Finally, for flows of KT and bi-KT geometries, the latter described in section \ref{bigeometries} below,   see \cite{ST1, ST2}.

\subsection{Hyper-K\"ahler Geometry with Torsion}\label{hkt}

An important special class of strong CYT manifolds are manifolds with a strong HKT structure (also called generalized hyper-K\"{a}hler manifolds).   The relation of strong HKT manifolds
to strong CYT manifolds roughly generalizes the relation between Calabi-Yau manifolds and hyper-K\"{a}hler manifolds to the case that a $B$-field is present.

A hypercomplex manifold is a manifold of dimension $D=2n=4k$  endowed with three integrable complex structures $I_r$, $r=1,2,3$ that satisfy
the algebra of imaginary unit quaternions:
$I_1^2=I_2^2=-1$, $I_1 I_2+I_2 I_1=0$, and $I_3=I_1 I_2$.
An HKT manifold (hyper-K\"{a}hler with torsion) is a hypercomplex manifold that is
 endowed with a metric $G$ that is hermitian with respect to all three complex structures,
and a connection $\hat\nabla$  with completely antisymmetric torsion $H$ such that  $G$ and all of the $I_r$ are covariantly constant with respect to $\hat\nabla$.
In particular, $M$ endowed with $G$ and any one of the $I_r$ (or more generally any real linear combination $\sum_r a_r I_r$ with $\sum_r a_r^2=1$) is a KT manifold.
Thus,  just as a hyper-K\"{a}hler manifold has many K\"{a}hler structures, an HKT manifold has many KT structures.   This  is the starting point of the twistor construction for HKT manifolds  \cite{HoweGP}.

A key point in the definition of an HKT manifold is  that, just as the same metric $G$ must be hermitian for all three complex structures, all three complex structures are required
to be covariantly constant for the same connection $\hat\nabla$, and hence  $H$ must be expressible as in (\ref{torsion}) with respect to each of the complex structure $I_r$. This is a rather restrictive condition.
 In particular,  $H$ must be of type $(2,1)\oplus (1,2)$ for each of $I_1,I_2,I_3$, since this is a general property of KT manifolds.

 As in the KT case, to each of the complex structures $I_r$ we associate a two-form $I_{r\, PQ}= G_{PR}I_r^R{}_Q.$   The quaternion relations satisfies by the $I_r$ imply, for example,
that $I_2+\i I_3$ is of type $(2,0)$ with respect to $I_1$ (and $I_2-\i I_3$ is of type $(0,2)$), along with obvious permutations of these statements.

One can define a Lee form $\hat\theta_r$ for each of the complex structures $I_r$ by the formula (\ref{leeform}). For HKT manifolds, all three Lee forms are equal \cite{IvanovGP}
\begin{eqnarray}
\hat\theta\equiv \hat\theta_1=\hat\theta_2=\hat\theta_3~.
\label{equallee}
  \end{eqnarray}
  One way to prove this is to use the fact that $H$ is of type $(2,1)\oplus (1,2)$ with respect to $I_1$ while $I_2+\i I_3$ is of type $(2,0)$.
  Hence
  \be\label{wilco} 0 = H_{PQR}\left(I_2+\i I_3\right)^{TP}\left(I_2+\i I_3\right)^{QR}. \ee
  The real part of this relation reads
  \be\label{realpart}   H_{PQR}I_2^{TP}I_2^{QR} =H_{PQR}I_3^{TP}I_3^{QR},\ee
  which says that $\h\theta_2=\h\theta_3$.   Similarly $\h\theta_1=\h\theta_2$.

It follows from covariant constancy of $G$ and of the $I_r$  that the holonomy of $\hat\nabla$ is contained in $Sp(k)\subset SU(n)\subset SO(N)$.
Accordingly, a strong HKT manifold is a special case of a strong CYT manifold.  In particular, therefore, as for any strong CYT manifold,
the curvature $\hat R$ of a strong HKT
manifold satisfies the condition (\ref{riccilee})  associated to scale invariance.

There are many examples of compact homogeneous strong HKT manifolds \cite{Spindel,OP}.  The simplest is $S^3\times S^1$, which we will discuss in section \ref{examples}.
Of course, any hyper-K\"{a}hler manifold is a strong HKT manifold.   The  compact smooth strong HKT manifolds that have been described in the literature
are locally the product of a hyper-K\"{a}hler manifold and a homogeneous strong HKT manifold.  If one drops the strongness condition $\d H=0$, then compact HKT manifolds can
be constructed as nontrivial fibrations of a homogeneous HKT manifold over a hyper-K\"{a}hler base, see e.g.  \cite{Verbitsky}.   If one relaxes the assumption of compactness, then many
more examples of strong HKT manifolds are known.

\subsection{Bi-KT and Bi-HKT Geometries}\label{bigeometries}

In the context of sigma-models, a KT or HKT structure can lead to enhanced supersymmetry.  If the target space $M$ of a sigma-model has a KT or HKT structure,
this can be used to enhance the right-moving supersymmetry of the sigma-model from $\N=1$ supersymmetry to $\N=2$ or $\N=4$, respectively.   Thus, $(0,1)$ supersymmetry
is enhanced to $(0,2)$ or $(0,4)$, and similarly $(1,1)$ supersymmetry is enhanced to $(1,2)$ or $(1,4)$.

To achieve extended supersymmetry for both left- and right-movers of the sigma-model, $M$ must be endowed with KT or HKT structures for both left-moving and right-moving
sigma-model modes.  Although it is possible to use the same KT or HKT structure for both left-movers and right-movers, this is not necessary.  The two structures can be chosen
independently \cite{JGCHMR}.    This leads to the idea of bi-KT and bi-HKT geometries.

  In the bi-KT case, $M$ is a manifold of dimension $D=2n$ that admits a pair of  KT structures with the same metric $G$ but  in general with different complex structures
  $\hat I$ and $\breve I$.   The two KT connections  $\hat\nabla$ and $\breve\nabla$ satisfy $\hat\nabla G=\breve\nabla G=0$,
  along with  $\hat\nabla \hat I=\breve\nabla\breve I=0$.  Since the sigma-model only has one $B$-field, the two connections are determined by the same
  three-form $H=\d B$.   However, since $B$ is odd under reversing the sigma-model worldsheet orientation (or exchanging left- and right-movers),
  the connections $\hat\nabla$ and $\breve\nabla$ actually are required to have equal and opposite torsion.  Thus $\hat\nabla$ is defined as in eqn. (\ref{condef}),
  and $\breve\nabla$ is defined by the same formula but with $H\to -H$.   The bi-KT structure of $M$ is said to be strong if $\d H=0$, as required in the construction
  of a sigma-model.

Strong bi-KT  geometries generically do not satisfy the scale invariance condition (\ref{scalecond}).  However, as in the KT case above, one can consider  strong bi-CYT geometries for which the holonomy of both $\hat\nabla$ and $\breve\nabla$ are restricted to lie in $SU(n)\subset U(n)\subset SO(D)$.  In such a case,
one has that $\hat R_{KL}=\hat\nabla_K \hat\theta_L$ and $\breve R_{KL}=\breve\nabla_K \breve \theta_L$, where  $\breve\theta_K=+\frac{1}{2} J^L{}_K H_{LPQ} J^{PQ}$. Moreover, the Bianchi identity (\ref{b2}) implies that $\breve R_{KL}=\hat R_{LK}$ and so one finds that
\begin{equation}
D_K (\hat\theta-\breve \theta)_L+D_L (\hat\theta-\breve \theta)_K=0~,~~~(\d \hat\theta+\d\breve \theta)_{KL}=H^P{}_{KL} (\hat\theta-\breve \theta)_P\,.
\label{cyt22}
\end{equation}
The first condition says that the dual of  $\hat\theta-\breve \theta$ is a Killing vector field that leaves the metric invariant, and the second condition says that
this vector field also leaves the $B$-field invariant up to a gauge transformation generated by the one-form $\h\theta+\breve\theta$.   So $\h\theta-\breve\theta$ generates
a symmetry of the sigma-model background.

A special case arises when the two CYT structures commute, i.e.
\begin{equation}
\hat I^K{}_L \breve I^L{}_P=\breve I^K{}_L \hat I^L{}_P\,,
\end{equation}
or more briefly $\hat I \breve I=\breve I \hat I$.
Taking the $D$ derivative of that  equation (where $D$ is the Riemannian connection),  and using the covariant constancy conditions $\hat\nabla \hat I=\breve \nabla \breve I=0$ as well as the integrability of $\hat I$ and $\breve I$, one can show that
 \be
 \hat\theta=\breve\theta\,.
 \label{hbtheta}
 \ee
Indeed, by taking the $D$-covariant derivative of $\hat I \breve I-\breve I \hat I=0$ and using $\hat\nabla \hat I=\breve \nabla \breve I=0$, one finds that
 \be
 H^F{}_{EA} \hat I_{FD} \breve I^D{}_B- H^F{}_{EB} \hat I_{AD} \breve I^D{}_F+ H^F{}_{ED} \hat I_{AF} \breve I^D{}_B+ H^F{}_{ED}  \hat I^D{}_B \breve I_{AF}=0\,.
 \ee
 Contracting this with $\hat I^{EA}$ leads to
 \be
 \breve I^D{}_B \hat I^F{}_D H_{FEA} \hat I^{EA}- H_{FDB} \breve I^{FD}+ H_{FED} \hat I^E{}_A \hat I^D{}_B \breve I^{AF}=0\,.
 \label{IIH}
 \ee
 As $H$ is of type $(2,1)\oplus (1,2)$  with respect to $\hat I$, it satisfies the condition
 \be
 H_{FED} \hat I^E{}_A \hat I^D{}_B+ H_{BED} \hat I^E{}_F \hat I^D{}_A+ H_{AED} \hat I^E{}_B \hat I^D{}_F- H_{FAB}=0\,.
 \ee
 Contracting this with $\breve I^{AF}$ and moving the last two terms of the above identity to the right hand side, one finds that
 \be
 2 H_{FED} \hat I^E{}_A \hat I^D{}_B \breve I^{AF}=- H_{BED} \hat I^E{}_F \hat I^D{}_A \breve I^{AF}- H_{AFB} \breve I^{AF}=0\,,
 \ee
 where the commutativity of the complex structures was again used in the last step. Substituting this into (\ref{IIH}) and multiplying with $\breve I^B{}_K$, we conclude that
 \be
 \hat I^F{}_K H_{FEA} \hat I^{EA}+ \breve I^F{}_K H_{FEA} \breve I^{EA}=0\,,
 \ee
  establishing (\ref{hbtheta}).

  In turn, the conditions (\ref{cyt22}) imply that $\theta=\hat\theta=\breve \theta$ is closed.  Therefore, at least locally one can introduce a dilaton $\Phi$ satisfying $\theta=-2\d\Phi$
  and thus satisfying the one-loop condition for sigma-model conformal invariance.
The condition $\hat I\breve I=\breve I \hat I$ was used in \cite{JGCHMR} as a necessary condition to describe a sigma-model in terms of chiral and twisted chiral superfields.
We will use this condition in section \ref{worldsuper} in discussing off-shell supersymmetry.

Bi-HKT geometries are defined similarly.  These admit two HKT structures with respect to the hypercomplex structures\footnote{From now on to simplify notation, we shall denote the HKT structures by only mentioning their associated hypercomplex structures.} $\hat I_r$ and $\breve I_r$, respectively, such that $\hat\nabla \h I_r=\breve \nabla \breve I_r=0$.   As
in the bi-KT case, $\h\nabla$ and $\breve\nabla$ are defined with equal and opposite torsion $H$.   The bi-HKT structure is called strong if $\d H=0$, as needed for application to
sigma-models (at least in Type II superstring theory).  On a bi-HKT manifold,
the Lee forms $\h\theta_r$ and $\breve\theta_r$ are independent of $r$ by virtue of eqn. (\ref{equallee}), so we denote them simply as $\h\theta$ and $\breve\theta$.
Since bi-HKT geometry is a specialization if bi-CYT geometry, $\h\theta$ and $\breve\theta$  are related as in (\ref{cyt22}). If in addition the two HKT structures commute, meaning
that  $\hat I_r \breve I_s=\breve I_s \hat I_r$ for all $r,s$, then it follows, specializing the bi-CYT result,  that $\hat\theta=\breve \theta$, so  we denote either of these as $\theta$. As
$\d\theta=0$, the  condition of one-loop conformal invariance can then be satisfied at least locally on $M$.

Similar considerations hold if $M$ admits an HKT structure  $\hat I_r$, $\hat\nabla \hat I_r=0$,  and a KT (CYT) structure $\breve I$, $\breve\nabla \breve I=0$.
If $\d H=0$, this pair of structures can be used to construct a sigma-model with $(2,4)$ supersymmetry.
 In the strong CYT-HKT case, the Lee forms $\hat\theta$ and $\breve \theta$ will satisfy (\ref{cyt22}).  If, in addition,  the strong HKT and CYT
 structures commute, $\hat I_r \breve I=\breve I \hat I_r$, then $\theta=\hat\theta=\breve \theta$ is a closed 1-form, and the condition of one-loop conformal invariance can be satisfied
 at least locally.

\subsection{Worldsheet Supersymmetry and Geometry}\label{worldsuper}

The  fields $X$ of a (1,1)-supersymmetric sigma-model are maps from a superspace  $\R^{2\vert1,1}$  with coordinates $(u,v\vert \vartheta^\pm)$ into a Riemannian manifold $M$ with metric $G$. Here   $(u,v)$ are  Grassmann  even light-cone coordinates  while  $\vartheta^\pm$ are the Grassmann  odd coordinates of $\R^{2\vert1,1}$.
The action of an (1,1)-supersymmetric sigma-model \cite{JGCHMR, Howesierra} with $G$ and $B$ couplings  written in terms of (1,1) superfields $X$ is
\be
S={1\over 4\pi \alpha'} \int_{\R^{2\vert1,1}} \d u\d v \d^2\vartheta\,\, (G+B)_{IJ} D_+ X^I D_- X^J\,,
\label{sigmaaction}
\ee
where  $D_\pm$ are  superspace derivatives that commute with the supersymmetry generators and satisfy
 $D_+^2=\i\partial_u$, $D_-^2=\i\partial_v$ and  $D_+ D_-+D_-D_+=0$. This action is manifestly invariant under (1,1) supersymmetry transformations.

If  the sigma-model target space admits either strong KT structure $\hat I$ ($\hat\nabla \hat I=0$), or strong HKT structure $\hat I_r$ ($\hat\nabla \hat I_r=0$), then the action (\ref{sigmaaction}) is invariant under   (1,2) or  (1,4)
supersymmetry transformations, respectively.  Following  \cite{JGCHMR,HoweGP1},  the additional supersymmetry transformations are given by
\be
\delta_{\hat I_r} X^K= \varepsilon^r \hat I_r^K{}_L D_+ X^L\,,
\label{hattransf}
\ee
with $r=1$ for the  KT case and $r=1,2,3$ for the HKT case, where $\varepsilon^r=\varepsilon^r(u, \vartheta^+)$ are  Grassmannian odd infinitesimal parameters.  These can depend on the superspace coordinates $(u, \vartheta^+)$, as classically the action is superconformally invariant.  The commutator of two such transformations is given by
\be
[\delta_{\hat I}, \delta'_{\hat I}] X^K=-2\i \varepsilon^r \varepsilon'^s \delta_{rs} \partial_u X^K+ (\varepsilon'^s D_+ \varepsilon^r \hat I_s^K{}_P \hat I_r^P{}_L-\varepsilon^r D_+\varepsilon'^s \hat I^K_r{}_P \hat I^P_s{}_L) D_+X^L\,,
\label{commut1}
\ee
where the integrability condition of the complex structures $\hat I_r$ has been used.\footnote{Vanishing of the Nijenhuis tensors of $\h I_r$ and $\h I_s$ was also used
to simplify their right hand side.} In the KT case, the second term in the right hand side of the commutator can be expressed as $(\varepsilon D_+\varepsilon'-\varepsilon' D_+\varepsilon) D_+X^K$, while in the HKT case it can be expressed as $\delta_{rs}(\varepsilon^r D_+\varepsilon'^s-\varepsilon'^sD_+\varepsilon^r ) D_+X^K- (\varepsilon^r D_+\varepsilon'^s+\varepsilon'^sD_+\varepsilon^r )\epsilon_{rs}{}^t \hat I_t{}^K{}_L D_+X^L$. In either case, the commutator (\ref{commut1}) closes to spacetime translations and supersymmetry transformations as expected.

Similarly, if the sigma-model target space admits either a strong bi-KT structure $(\hat I, \breve I)$ or a strong bi-HKT structure $(\hat I_r, \breve I_r)$, then the action (\ref{sigmaaction}) is invariant under (2,2) or (4,4) supersymmetry transformations, respectively. The additional supersymmetry transformations are given by
\be
\delta_{\hat I} X^K= \hat \varepsilon^r \hat I_r^K{}_L D_+ X^L~,~~~\delta_{\breve I} X^K= \breve \varepsilon^r \breve I_r^K{}_L D_- X^L\,,
\label{hattransf2}
\ee
where the $\delta_{\hat I}$ transformations are as in (\ref{hattransf}) and the infinitesimal parameters $\breve \varepsilon^r$ of the $\delta_{\breve I}$
transformations may depend on the worldsheet superspace as $\breve \varepsilon^r=\breve \varepsilon^r(v, \vartheta^-)$. Again this is because the action
(\ref{sigmaaction}) is classically invariant  under  superconformal transformations.
The commutators $[\delta_{\hat I}, \delta'_{\hat I}]$  are given as in (\ref{commut1}).  The commutator
$[\delta_{\breve I}, \delta'_{\breve I}]$ is also given as in  (\ref{commut1}) after replacing the parameters $\hat \varepsilon$
and $\hat \varepsilon'$ with $\breve \varepsilon$ and $\breve \varepsilon'$, respectively, and the $D_+$ and $\partial_u$
derivatives on the fields $X$ with $D_-$ and $\partial_v$. So far, the commutators close to spacetime translations and worldsheet supersymmetry transformations.

The remaining commutator, upon using $\hat\nabla \hat I_r=\breve \nabla \breve I_s=0$,  can be arranged as
\be
[\delta_{\hat I}, \delta_{\breve I}] X^K= \hat \varepsilon^r \breve \varepsilon^s (\hat I_r^K{}_P \breve I^P_s{}_L-\breve I^K_s{}_P \hat I_r^P{}_L) \hat\nabla_+D_- X^L\,,
\ee
where $r=s=1$ for bi-KT geometries and $r,s=1,2,3$ for bi-HKT geometries.
Clearly, the commutator fails to  close off-shell unless $\hat I_r$ and $\breve I_s$ commute.  If $\hat I_r$ and $\breve I_s$ do commute, the
algebra of transformations (\ref{hattransf2}) closes off-shell to  (2,2) or (4,4) supersymmetry. A similar analysis applies for $(2,4)$ or $(4,2)$ supersymmetry.

Provided that all conditions  for the off-shell closure of worldsheet supersymmetry transformations  of a $(p,q)$ supersymmetric
sigma-model are met, one can construct a (standard) $(p,q)$ superfield description of the theories. The $(p,q)$ superfields $X$ are
maps for the $\R^{2\vert p,q}$ superspace with coordinates $(u,v\vert \vartheta^{+0}, \vartheta^{+r}, \vartheta^{-0}, \vartheta^{-s})$,
$r=1,\dots, q-1$ and $s=1, \dots, p-1$, into the sigma-model target manifold $M$ that satisfy the constraints \cite{HoweGP1, HoweGP2}
\be
D_{+r}X^K=\hat I^K_r{}_L D_{+0} X^L~,~~~D_{-s}X^K=\breve I^K_s{}_L D_{-0} X^L\,.
\label{pqsuper}
\ee
The conditions we have found above for the off-shell closure of the supersymmetry algebra arise as integrability conditions for these constraint equations.   A superfield action is
\be
S={1\over 4\pi \alpha'} \int_{\R^{2\vert1,1}} \d u \d v \d^2\vartheta^{0}\,\, (G+B)_{IJ} D_{+0} X^I D_{-0} X^J\,,
\label{sigmaaction2}
\ee
where the superfields $X$ satisfy the constraint (\ref{pqsuper}).
 It can be shown that this  action is invariant under all $(p,q)$ supersymmetry transformations.

Note that  the non-linear constraints (\ref{pqsuper})  on the fields can be linearised after an appropriate choice of coordinates on $M$. For example the conditions on (0,2) and (1,2) superfields can be linearised using complex coordinates on $M$ \cite{JGCHMR, Hullwitten}. However, this is not always the case. In some models additional conditions are needed on $M$, which do not arise as integrability conditions of (\ref{pqsuper}). For example the linearisation of the constraints \cite{HoweGP2} for the (0,4) and (1,4) superfields  requires the vanishing of the curvature of the Obata connection  of the hypercomplex structure.    (The Obata connection on a hypercomplex manifold is the unique torsion-free connection for which the
complex structures are covariantly constant; it is not metric compatible.)

\subsection{Examples}\label{examples}

 Here, we shall describe in detail some examples of HKT manifolds that are useful for illustrating the relation of scale and conformal invariance in sigma-models.  We consider
 first homogenous HKT structures on $S^3 \times S^1$, and then we consider some deformations of (portions of) this space.
We endow\footnote{There is an extensive literature on different aspects of the  the $S^3\times S^1$ model, for example \cite{Sevrin}-\cite{ADRS}. See also \cite{Ivanov} for a harmonic superspace treatment of (4,4) supersymmetric sigma models that includes an application to the $S^3\times S^1$ model.
$S^3\times S^1$ is an example in which the Obata connection, mentioned at the end of section \ref{worldsuper}, has vanishing curvature.
The  superfield constraints  (\ref{pqsuper}) can be linearised in this model  \cite{HoweGP2} and the geometry can be written in terms of a quaternionic coordinate $Q$:  $\d s^2= (\bar Q Q)^{-1} \d\bar Q \d Q$. In these coordinates, the commuting bi-HKT structure $(\hat I^-, \breve I^+)$, given later,  is associated with left and right multiplication on $Q$ with the quaternionic imaginary units.  The model constructed  with these commuting hypercomplex structures can be described with one chiral supermultiplet and
one twisted chiral supermultiplet \cite{RocekSS}.   Twisted chiral supermultiplets were introduced in \cite{JGCHMR}.}   $S^3\times S^1$ with an obvious homogeneous metric:
\be\label{prodmet}\d s^2=4\d\Omega^2 + \d \tau^2, \ee
where $\d\Omega^2$ is a round metric on $S^3$ of  radius 1, and $\tau$ is a periodic variable with an arbitrary period $T$, $\tau\cong t+T$.  A factor of 4 was included
in eqn. (\ref{prodmet}) to avoid factors of 2 later; of course, the metric in (\ref{prodmet}) could be rescaled by any constant factor without affecting the HKT condition.

We recall that $S^3$ can be identified as the $SU(2)$ group manifold, and that the round metric on $S^3$ is invariant under the left and right action of $SU(2)$ on itself.
Similarly, we can view $S^3 \times S^1$ as the group manifold $K=SU(2)\times U(1)$.   On this group manifold, we can pick orthonormal bases of left- and right-invariant
one-forms.  The form $L^0=R^0=\d\tau$ is both left- and right-invariant, as $U(1)$ is abelian.
On $S^3$ we can pick a basis\footnote{\label{details} Explicitly, parametrize  $S^3$ by functions $y_0,\cdots, y_3$ satisfying $\sum_{i=0}^3 y_i^2=1$, and set $L^1= 2(y_0 \d y_1 -y_1\d y_0
+ y_2\d y_3-y_3\d y_2)$, with $L^2$ and $L^3$ differing from this by cyclic permutations of indices $1,2,3$.   Then $\d L^1= L^2\wedge L^3$ (along with cyclic permutations of this
statement) and
$L^1\otimes L^1+L^2\otimes L^2+L^3\otimes L^3=4\sum_i \d y_i^2=4\d\Omega^2$. Because the $L^i$ are left-invariant, it suffices to verify these statements at the point $p$ defined by
$(y_0,y_1,y_2,y_3)=(1,0,0,0)$.  Similarly, one can take $R^1=2(y_0 \d y_1 -y_1\d y_0
-y_2\d y_3+y_3\d y_2)$, with $R^2$ and $R^3$ obtained by cyclic permutations of indices $1,2,3$, leading to $\d R^1=-R^2\wedge R^3$, and cyclic permutations.
The $L$'s and $R$'s are respectively left-invariant and right-invariant as they were constructed using antisymmetric $4\times 4$ matrices that are respectively self-dual or
anti-self-dual.
 As the $L$'s and $R$'s are both  orthonormal bases, $L^1\wedge L^2\wedge L^3$
and $R^1\wedge R^2\wedge R^3$  both equal the volume form of $S^3$,  up to sign,  and hence must be equal up to sign.   A short calculation at the point $p$ confirms that they are equal.}
 of left-invariant one-forms $L^1,L^2,L^3$, normalized so that
\be\label{leftinv}\d L^1=L^2\wedge L^3,\ee
and cyclic permutations of this statement.
  Similarly, we can pick a basis of right-invariant one-forms $R^1,R^2,R^3$, satisfying
\be\label{rightinv}\d R^1=-R^2\wedge R^3\,,\ee
and cyclic permutations. Then $L^a$, $a=0,\cdots, 3$ is a basis of left-invariant one-forms, and $R^a$, $a=0,\cdots, 3$, is a basis of right-invariant ones.
These bases are orthonormal, meaning that the line element defined in eqn. (\ref{prodmet}) satisfies
\be\label{metsum}\d s^2=\sum_{a=0}^3 L_a\otimes L_a=\sum_{a=0}^3 R_a\otimes R_a\,. \ee
We can define a connection $\h\nabla$ on the tangent bundle of $S^3\times S^1$ by saying that the one-forms $L^a$ are covariantly constant,
and another connection $\breve\nabla$ by saying that the one-forms $R^a$ are covariantly constant.   From eqn. (\ref{condef}), it follows that for any one-form $V$ and metric compatible connection $\h\nabla$ with completely antisymmetric torsion $H$,
one has $\h \nabla_I V_J=\nabla_I V_J - \frac{1}{2}H_{IJK}V^K$.  Consequently, if $\h\nabla V=0$, then $\nabla_I V_J-\nabla_J V_I=H_{IJK}V^K.$  Taking $V=L^1$, and using (\ref{leftinv}),
we find that the torsion $H$ of $\h\nabla$ must equal the volume form $L^1\wedge L^2\wedge L^3$ of $S^3$.     By the same reasoning, the torsion of $\breve\nabla$
is $-R^1\wedge R^2\wedge R^3$.     In fact, $L^1\wedge L^2\wedge L^3=R^1\wedge R^2\wedge R^3$ (see footnote \ref{details}), so the connections $\h\nabla $ and
$\breve\nabla$ have equal and opposite torsion.

To find an HKT structure on $S^3\times S^1$ with connection $\h\nabla$, what we still need is to describe a hypercomplex structure such that the complex
structures   are covariantly constant with respect to $\h\nabla$, and the metric is of type $(1,1)$ with respect to each complex structure.   There actually are two natural left-invariant HKT structures,  compatible
with the same connection $\h \nabla$, and  differing
by a choice of orientation of $S^1$, and similarly there are two natural right-invariant HKT structures,  differing in a similar way and compatible with the connection $\breve\nabla$.

For one left-invariant HKT structure,  we can define a complex structure $\h I_1^+$ by
\be\label{zimbo}\h I_1^+(L^0)=L^1\,,~~\h I_1^+(L^1)=-L^0\,,~~~\h I_1^+(L^2)=L^3\,,~~~\h I_1^+(L^3)=-L^2\,. \ee
The resulting complex structure can be described as follows.   There is a left-invariant Hopf fibration $S^3\to S^2$ with fiber $S^1$.   Taking the product with another $S^1$,
we get a fibration $S^3\times S^1\to S^2$ with fiber $S^1\times S^1$.   Both the base space of this fibration $S^2\cong \mathbb{CP}^1$ and the fiber $S^1\times S^1=T^2$ are
complex maifolds.  The structure group
of the fibration is the $U(1)$ group of rotations of the first factor of $T^2=S^1\times S^1$, which acts holomorphically on $T^2$. So the total space of the fibration,  namely  $S^3\times S^1$,
is a complex manifold.
The other two complex structures making up the hypercomplex structure are obtained from eqn. (\ref{zimbo}) by cyclic permutations of indices $1,2,3$.   They do obey the expected quaternion relations
including $\h I_1^+\h I_2^+=+\h I_3^+$.

The corresponding hermitian forms are
\be\label{timbo} \h I_1=L^0\wedge L^1+L^2\wedge L^3\,,\ee
along with similar formulas obtained by cyclic permutations of indices.   These hermitian forms or equivalently the corresponding complex structures are all covariantly constant for the connection
$\h\nabla$, since the $L$'s are covariantly constant.   So this defines an HKT structure, which is strong as $\d H=0$.

To get a second HKT structure also compatible with the same connection, we can simply change coordinates in $M$ by $\tau\to -\tau$, resulting in $L^0\to -L^0$.  However,  if we merely reverse the
sign of $L^0$ in eqn. (\ref{zimbo}) and the corresponding cyclically permuted formulas,  we will get three complex structures that will satisfy $I_1 I_2=-I_3$,  not the standard form of the quaternion
relations.   We can compensate for this by reversing the sign of all the $I$'s.   Thus we can define a second HKT structure
via the complex structure
\be\label{limbo}\h I_1^-(L^0)=L^1\,,~~\h I_1^-(L^1)=-L^0\,,~~~\h I_1^-(L^2)=-L^3\,,~~~\h I_1^-(L^3)=+L^2\,,\ee
or equivalently the hermitian form
\be\label{timbox} \h I^-_1=L^0\wedge L^1-L^2\wedge L^3\,,\ee
along with their cyclically permuted relatives.

We can also define a pair of right-invariant HKT structures using the same formulas but with the $L$'s replaced by the $R$'s.  Thus the complex structures
are
\be\label{kimbo}\breve I^\pm_1(R^0)=R^1,~~~\breve I^\pm_1(R^1)=-R^0,~~ \breve I^\pm_1(R^2)=\pm R^3,~~\breve I^\pm_1 (R^3)=\mp R^2\,,\ee
plus cyclic permutations, and equivalently the hermitian forms are
\be\label{vimbo}  \breve I^\pm_1 = R^0\wedge R^1\pm R^2\wedge R^3\,,\ee
plus permutations.

To make a bi-HKT geometry,  we can take either of the two left-invariant structures and pair it with either of the two right-invariant structures.   However,  since a change of coordinates $\tau\to -\tau$
exchanges the two left-invariant structures and likewise exchanges the two right-invariant structures,  there are only two essentially different cases: we can pair the hypercomplex structure
defined by $\h I^+_r$
with the one defined by either $\breve I^+_s$ or $\breve I^-_s$.    The two choices differ by whether or not the two hypercomplex structures commute.   Because all the hypercomplex structures considered are
left- or right-invariant,  to decide if two of them commute,  it suffices to check that at a single point in $S^1\times S^3$,  for instance the product of any point in $S^1$ with the point $p\in S^3$ defined in
 footnote \ref{details}.   In the tangent space at this point,  selfdual matrices commute with anti-selfdual matrices,  but selfdual or antiselfdual matrices do not commute with matrices of the same type.
 Moreover,  $\h I^+_r$ and $\breve I^+_s$ act in the tangent space at the given point by selfdual matrices, but $\h I^-_r$ and $\breve  I^-_s$ act by anti-selfdual matrices.  So to get a bi-HKT
 structure with commuting hypercomplex structures,  we should pair $\h I^+_r$ with $\breve I^-_s$ or $\h I^-_r$ with $\breve I^+_s$.    The other pairings give non-commuting hypercomplex structures.

  This complete our description of the basic $S^3\times S^1$ example.
In what follows, we will consider two types of modification of this example.
The two constructions differ by whether the two hypercomplex structures  commute.
First we consider certain four-dimensional bi-HKT geometries
\cite{GP, Tod} that admit a triholomorphic Killing vector field $V$ that leaves everything  invariant, i.e. ${\cal L}_VG={\cal L}_V \hat I_r={\cal L}_V H=0$ (where $\cal L$ is the Lie derivative).     The metric $G$
 and torsion $H$ of the example constructed in \cite{GP}  are described by
\begin{equation}
\d s^2 = W^{-1} (\d\tau+\omega)^2+ W \d s^2_{S^3}~,~~~H= W \d\mathrm{vol}(S^3)\,,
\label{trihkt}
\end{equation}
where $\d s^2_{S^3}$ is the  line element of a round three-sphere normalized as before in (\ref{metsum}) and $\d\mathrm{vol}(S^3)$ is the volume form of the sphere.    The function $W$ and one-form $\omega$ are
assumed to be invariant under the symmetry generated by the Killing vector field $V=\frac{\partial}{\partial \tau}$,   and to be
related by
$\star_3 \d\omega= \d W$, where  $\star_3$  is the Hodge star operator of $S^3$.
This implies,  in particular, that $W$ must be a harmonic function on $S^3$.  Of course,  $W$ must be positive in order for the metric (\ref{trihkt}) to be well-defined.
 An everywhere smooth harmonic function would have to be constant, leading back to the  the original $S^3\times S^1$ geometry.
However,  there certainly exist functions $W$ that are harmonic and positive on an open set $M\subset S^3\times S^1$.   Restricting to such an open set $M$,  this construction gives
new (incomplete) strong HKT geometries.

It is convenient to use the same left-invariant one-forms  $L^r$ as before.  An orthonormal co-frame for the metric $G$  is
 \be
 E^0=W^{-\frac{1}{2}} (\d\tau+\omega)\,,~~~E^r=W^{\frac{1}{2}} L^r\,.
 \label{coframeE}
 \ee
In this frame,  the Hermitian forms of the hypercomplex structure are
\be
\hat I_r^{-}= E^0\wedge E^r- \frac{1}{2} \epsilon^r{}_{st} E^s\wedge E^t\,.
\label{triherm}
\ee
 The Lee form is $\hat\theta=W^{-1} (\d\tau+\omega)$.  The dual vector field is the Killing vector field  $V$,  which is evidently triholomorphic as it generates a symmetry that leaves invariant
  the hermitian forms as well as the metric.

The same  geometry (\ref{trihkt}) admits another HKT structure $\breve I_r^{-}$ with equal and opposite torsion  \cite{Tod}.  The Hermitian forms are
 \be
 \breve I_r^{-}=\tilde E^0\wedge \tilde E^r- \frac{1}{2} \epsilon^r{}_{st} \tilde E^s\wedge \tilde E^t\,,
 \label{triherm2}
  \ee
where \be\label{frames}\tilde E^0=W^{-\frac{1}{2}} (\d\tau+\omega), ~~\tilde E^r=W^{\frac{1}{2}} R^r\,,\ee
with $R^r$ as before.    The hypercomplex structures  $\breve I_r^{-}$ and  $\hat I_r^{-}$ do not commute,  since indeed they do not commute in the special case that $W$ is constant.

A different deformation of a portion of $S^3 \times S^1$
 preserves a strong bi-HKT structure in which the two hypercomplex structures commute.   Before explaining this, let us note than in four dimensions,
any hyperhermitian metric on a hypercomplex manifold $M$ is an HKT metric  (see Lemma 1 in \cite{BaSw}).   It is also true that any two hyperhermitian metrics on a hypercomplex  four manifold differ only
by a Weyl rescaling.   And since  the space of four-forms in four dimensions is one-dimensional,  the strong HKT equation $\d H=0$ is a single differential equation for the Weyl factor;  this turns out to be a Laplace-like equation that always has local solutions,  though it may lack nonsingular global solutions.

One can recover some well-known solutions by implementing this idea for $\R^4$ with its standard hypercomplex structure.   In fact,  $\R^4$ has two standard and commuting hypercomplex structures,  with the hermitian
forms being selfdual  or anti-selfdual two-forms.    The standard flat metric  is hyperhermitian for these hypercomplex structures,  and any other hyperhermitian metric is
conformally equivalent to this one.   So we consider the metric
\be\label{conmet}G_{IJ}=e^{2\Phi}\delta_{IJ}\,,  \ee
with a scalar function $\Phi$.   A short calculation shows that this metric is bi-HKT,  with $H=-\frac{1}{2}\star \d e^{2\Phi}$ (here $\star$ is the Hodge star defined with respect to the flat metric
$\delta_{IJ}$).       The strong HKT condition $\d H=0$ becomes $\Delta^2 e^{2\Phi}=0$,  where $\Delta^2=\sum_{I=1}^4\partial_I^2$ is the Laplacian with respect to the flat metric.
The Lee form is $\theta= -2\d\Phi$.    It is closed,  as is the case for any bi-HKT geometry with commuting hypercomplex structures,  according to  the general analysis in section \ref{bigeometries}.

The solution associated to the NS5-brane \cite{Strominger, Duff,CHS}  is\footnote{An arbitrary factor of 4  is included here for convenience.}
\be\label{ns5brane}e^{2\Phi}=1+\frac{4}{\vec X^2}\,,~~~~\vec X^2=\sum_I (X^I)^2\,. \ee
 In the near horizon region,  one drops the constant term and replaces this with
 \be\label{nearh} e^{2\Phi}=\frac{4}{\vec X^2}\,.\ee
 With or without the constant term,  this gives a solution that is smooth away from the point $X=0$.    The one-loop condition for conformal invariance is satisfied, with $\Phi$ as the dilaton.
 The central charge is $\h c=4$.

 However,  we can do the following.   The metric in the near horizon geometry is described by the line element
 \be\label{nearline}\d s^2=4\frac {\d \vec X^2}{\vec X^2}\,.\ee
 This is scale-invariant.   So we can take the quotient by the group $\Z$ acting via $X\to e^{T/2} X$ (for any chosen constant $T>0$).  This will give a solution with target space $S^3\times S^1$
 where $S^3$ has radius 2 and $S^1$ has circumference $T$.   We have recovered the bi-HKT geometry $S^3\times S^1$ with commuting hypercomplex structures that was described earlier.
 A single-valued dilaton does not exist,  since $\Phi$ is not invariant  under the rescaling of $X$.   Nevertheless,  as explained in our earlier discussion of $S^3 \times S^1$,  since the Lee form
 is covariantly constant for the connection with torsion (in either of the two hypercomplex structures),  the model satisfies the condition for conformal invariance with constant dilaton,  leading to
 a conformal field theory with $\h c<4$.

 At the cost of introducing some singularities,  we can easily modify the construction so that the Lee form will no longer be covariantly constant for the connection with torsion.
 To do so,  we simply modify $\Phi$,  preserving the fact that $e^{2\Phi}$ satisfies the Laplace equation and is invariant under the rescaling $X\to e^{T/2}X$.   For example,  we can pick two generic
 points $a,b\in \R^4$ and a small parameter $\epsilon>0$ and take
 \be\label{earline}e^{2\Phi}=\frac{4}{\vec X^2}+\epsilon\sum_{n\in \Z}\left( \frac{1}{|\vec X - e^{n T/2} a|^2} -\frac{1}{|\vec X-e^{n T/2}b|^2}\right)\,. \ee
 The sum over $n$ converges exponentially fast for $n\to \pm \infty$; for $n\to -\infty$, this depends on a cancellation between the two series with $a$ and $b$.  To achieve
 this cancellation,  one of the series must occur with a negative coefficient, as in eqn.  (\ref{earline}).     The solution can only be defined on the open set  $M\subset S^3 \times S^1$ on which
 $e^{2\Phi}>0$.    For small $\epsilon$,  $M$ can be described approximately as the complement of a small ball around a point $b\in S^3 \times S^1$.   In particular,  the fundamental
 group of $M$ coincides with the fundamental group of $S^3 \times S^1$,  namely $\Z$.
 It remains true that $\Phi$ cannot be defined as a single-valued function on $M$.

\subsection{Scale and Conformal invariance Revisited}\label{scalerevisited}

A notable fact about strong HKT geometry and related geometries is that, generically, extended worldsheet supersymmetry with a generalized Calabi-Yau
condition ensures  that a sigma-model satisfies the one-loop condition for scale-invariance but does not guarantee the one-loop condition
for conformal invariance \cite{Buscher,Hull,HullTownsend}.  As we have reviewed,  the obstruction involves the
 Lee form $\theta$.  The one-loop condition for scale-invariance is always satisfied, with the vector field $V$ of eqn. (\ref{scalecond}) being the dual of $\theta$, while the
 generator $\Lambda$ of a $B$-field gauge transformation is simply $\theta$.
Conformal invariance requires that $\theta$ should be the sum of two terms:  the gradient of a scalar function,  and  a one-form that is covariantly constant for the connection
$\h\nabla$.  Such a one-form is
the dual of a Killing vector field $V$  that satisfies
\be\label{satisfactory}
V^I H_{IJK}+\partial_J V_K-\partial_K V_J=0\,,\ee
as well as the Killing vector condition $D_I V_J+D_J V_I=0$.    The gradient of a scalar function can be interpreted as the sigma-model dilaton, and a one-form that is covariantly
constant for $\h\nabla$  does not actually contribute in the condition  $\h R_{KL}=\h\nabla_K\theta_L$ (eqn. (\ref{riccilee})) for global scale invariance.

The one-loop condition for conformal invariance may be obstructed at two levels.    It may be impossible even locally to write $\theta$ as the sum of
an exact one-form and a covariantly constant one-form,  or this may be possible locally but not globally.

In the case of a generic strong CYT or strong HKT geometry, we have found nothing that would indicate that $\theta$ is locally the sum of an exact one-form and a covariantly constant one-form.
Hence it is reasonable to expect that generically there will be a local obstruction to the one-loop condition for conformal invariance.
 On the other hand,  in the important special case of a strong bi-HKT geometry with commuting hypercomplex structures,  we proved in
section \ref{bigeometries} that $\d\theta=0$.   Therefore, at least locally, $\theta=-2\d\Phi$ for some scalar function $\Phi$,  and the condition of conformal invariance can be satisfied
at least locally,  but there might be a  global obstruction.

But have we overlooked some additional constraints that follow from the geometries in question?
For example, it took a fairly elaborate argument to show in section \ref{bigeometries} that $\d\theta=0$ for a certain class of geometries.   Might  a more complete
analysis  place   constraints beyond those that we know?   In this section, we will use the examples described in section \ref{examples} to show that the
obstructions to conformal invariance that are allowed by the conditions we know  actually can occur.

First we consider the homogeous bi-HKT model with target space $S^3\times S^1$.   This is an example in which the one-loop condition for conformal invariance is
{\it not} obstructed.   (Indeed, the sigma-model with this target space is a conformally-invariant WZW model.)  For $M=S^3\times S^1$, we found the Lee form to be
$\theta=\d\tau$.  This is covariantly constant for the connection with torsion.   Hence the condition for one-loop (and even exact) conformal invariance can be satisfied,
with a constant dilaton that does not contribute to the beta functions.

The model obtained this way has $\hat c<4$ (since  the WZW model with target $S^3\cong SU(2)$ has $\h c<3$).   Something else that we can do with the same model
is to replace $S^1$ with its universal cover $\R$.    On the universal cover, the closed one-form $\theta=\d\tau$ becomes exact, so we can achieve conformal invariance
by introducing a linear dilaton, proportional to $\tau$.   This gives a different model with a noncompact target space and $\h c=4$.

In section \ref{examples}, we discussed two ways to make a singular perturbation of the HKT manifold $S^3\times S^1$.  Both of these constructions led to strong bi-HKT geometries,
and therefore, at least in perturbation theory,\footnote{The reason that we say ``at least in perturbation theory'' is that in both cases the perturbations were singular
and the perturbed target space is singular and noncompact.  Hence it is not clear that the corresponding sigma-models make sense nonperturbatively.} to models with $(4,4)$ supersymmetry.
The two examples differ by whether or not the two hypercomplex structures commute.

 In eqns. (\ref{trihkt})-(\ref{triherm2}), we described a strong bi-HKT deformation of an open
set  $M\subset S^3\times S^1$  in which the two hypercomplex structures do not commute.    In this example, the one-loop condition for scale-invariance cannot be satisfied even locally. The Lee form of the connection $\h\nabla$ is
$\h\theta= W^{-1}(\d\tau+\omega)$, which actually is dual to the Killing vector field $V=\frac{\partial}{\partial \tau}$.   There is no Killing vector field that is covariantly
constant with respect to $\h\nabla$ (for generic $W$, $V$ is the only Killing vector field, and it is not covariantly constant).   So to satisfy the one-loop condition
for conformal invariance, we would need a scalar function $\phi$ such that $\h\theta=2\d\phi$.   Such a function does not exist even locally, as $\d\h\theta\not=0$.

On the other hand, in  eqn.  (\ref{earline}), we described a strong bi-HKT deformation of an open set  $M\subset S^3\times S^1$, with commuting hypercomplex
structures.    Generically, $M$ has no Killing vector
fields, so to satisfy the one-loop condition for conformal invariance, we need $\h\theta=2\d\phi$ for some function $\phi$.
This is possible locally, since, as follows from a general argument explained in section \ref{bigeometries}, in this model $\h\theta$ is closed.
 But if the parameter $\epsilon$ in eqn.  (\ref{earline})  is
sufficiently small  so that $M$ inherits the fundamental group of $S^3\times S^1$, then globally $\h\theta$ is not exact and the condition of one-loop conformal
invariance cannot be satisfied globally.

Thus, at least for these particular questions, the general  behavior is no better than is predicted by the constraints we know.  As originally suggested in
\cite{Polchinski}, the examples that show a discrepancy between the conditions for scale-invariance  and for conformal-invariance involve non-compact (and incomplete) target spaces.
This is in accord with the general argument in \cite{Polchinski} concerning two-dimensional field theories with a discrete spectrum of operator dimensions, as well as the explicit
sigma-model argument that we described in section \ref{scaleconf}.

\appendix

\section{ Bianchi Identities} \label{ap-a}

As we have used throughout the paper the Bianchi identities  of connections with torsion, we collect them here for completeness.
The Riemann tensor $\h R_{LN}{}^K{}_P$ of a connection $\h \nabla$  with torsion $H$ is defined by
\begin{equation}\label{riemann}
\hat\nabla_L \hat\nabla_N V^K-\hat\nabla_N \hat\nabla_L V^K=\hat R_{LN}{}^K{}_P X^P- H^P{}_{LN} \hat\nabla_P V^K\,,
\end{equation}
for any vector field $V$.
Assuming that $\d H=0$,
the first Bianchi identities are
\begin{equation}
\hat {R}_{K[LPQ]}=-\frac{1}{3} \hat\nabla_K H_{LPQ}\,,
\label{b1}
\end{equation}
and
\begin{equation}
\hat R_{KLPQ}=\breve {R}_{PQKL}\,,
\label{b2}
\end{equation}
where $\hat R_{KL,PQ}$ is the curvature of the metric-compatible connection $\hat\nabla$ with torsion $H$ and  $\breve R_{KL,PQ}$ is the curvature of the metric-compatible
connection $\breve\nabla$ with torsion $-H$.

  The second Bianchi identity gives
\begin{eqnarray}
&&\hat\nabla_K\hat R_{LP QW}+\hat\nabla_P\hat R_{KL QW}+\hat\nabla_L\hat R_{PK QW}=H^Z{}_{KL} \hat R_{PZQW}
\cr
&&\qquad\qquad\qquad\qquad +H^Z{}_{PK} \hat R_{LZQW}+H^Z{}_{LP} \hat R_{KZQW}\,.
\label{b3}
\end{eqnarray}
After contracting with the metric, this implies that
\begin{equation}
\hat\nabla^Q\hat R_{LP QW}+\hat\nabla_P\hat R_{LW}-\hat\nabla_L\hat R_{PW}=H^{ZQ}{}_{L} \hat R_{PZQW}-H^{ZQ}{}_{P} \hat R_{LZQW}+H^Q{}_{LP} \hat R_{QW}\,.
\label{b4}
\end{equation}
A second contraction and use of eqn. (\ref{b1}) gives
\begin{equation}
\hat\nabla^L\hat R_{KL}-{1\over2}D_K\hat R={1\over12} D_K H^2-H^{LP}{}_K \hat R_{LP}\,.
\label{b5}
\end{equation}
In these formulas, $H^2=H_{KLP} H^{KLP}$, $\hat R_{NP}=\hat R_{LN,}{}^L{}_P$ is the Ricci tensor, and $\hat R=G^{NP} \hat R_{NP}$ is the Ricci scalar.

\section{Another Approach To Scale-Invariance and Conformal-Invariance}\label{another}

In section \ref{scaleconf}, we explained a proof of the relation between scale-invariance and conformal invariance following the logic of \cite{Perelman,OSW}.
In this appendix, we will instead describe another attempt that does not quite succeed, but curiously
would succeed if one artificially adds to the metric beta function $\beta_G$ a cosmological
constant term $-\lambda G_{IJ}$ with $\lambda\not=0$.   The interpretation of this is not clear in sigma-models (though inclusion of this term is important
in the context of Ricci flow, as already remarked in footnote \ref{alsoimportant}).

We recall that the metric compatible connection  $\hat\nabla$  with torsion $H$ satisfies $\hat\nabla_I V^J=D_I V^J+\frac{1}{2} H^J{}_{IK} V^K$.
We will consider the scale-invariance condition in the form in which it naturally arises in HKT geometry, but with an extra term $-\lambda G_{IJ}$ added on the left
hand side:
\be
\hat R_{IJ}-\lambda G_{IJ}=\hat\nabla_I V_J\,.
\label{scalecond2}
\ee
Here $\hat R_{IJ}=\hat R_{KI}{}^K{}_J$ is the Ricci tensor of $\hat \nabla$.
The conformal invariance condition, also with the extra term added, is
\be
\hat R_{IJ}-\lambda G_{IJ}+2\hat\nabla_I \partial_J \Phi=0\,.
\label{conformalcond2}
\ee
It turns out  that if $\lambda\not=0$, a solution of eqn. (\ref{scalecond2}) on a compact manifold $M$ is actually, for suitable $\Phi$, a solution of eqn. (\ref{conformalcond2}).

We will first discuss a direct attempt to imitate Bourguignon's argument \cite{Bourg}.
Using (\ref{scalecond2}) and taking the divergence of the Bianchi identity (\ref{b5}) with respect to  $\hat\nabla$, one finds,  after some computation, that
\begin{equation}
\Delta^2 \left(-\frac{1}{2} \hat R+\frac{1}{12} H^2\right)=V^I D_I \left(\frac{1}{2} \hat R+\frac{1}{12} H^2\right)+ \hat R_{IJ} \hat R^{IJ}-\lambda \hat R\,,
\label{maxprin}
\end{equation}
where $\hat R$ is the Ricci scalar of $\hat\nabla$ and $\Delta^2=G^{IJ}D_ID_J$ is the Laplacian.
For  $H=\lambda=0$, this formula reduces to that of  Bourguignon.
Bourguignon's argument was based on applying the maximum principle to this formula.\footnote{Assume that $\lambda = H=0$,  so that in particular there is no
distinction between $R_{IJ}$ and $\h R_{IJ}$.    By taking the trace in eqn. (\ref{scalecond2}) and integrating over $M$,  one learns that the
average value of $R$ is zero.     In eqn. (\ref{maxprin}),  at a point at which $R$ achieves its minimum,  the left hand side is non-positive and the right hand side is non-negative,  so both must vanish.   At this point,  therefore,  $R_{IJ}=0$,  implying that $R=0$.   Hence the minimum value of $R$ is zero, and as the average value is also zero,  it follows that $R$ vanishes identically.   But then eqn.  (\ref{maxprin}) implies
that $R_{IJ}=0$,  so that the condition of conformal invariance is satisfied with constant dilaton. For $\lambda=0$, the same argument  implies that $\h R_{IJ}=0$ provided one of the following three conditions is satisfied: $\hat R=0$, $H^2$ constant or $R$ constant.} The $H^2$ terms do not seem to enter in a convenient way for this argument.

However,  us try to solve the conformal invariance condition (\ref{conformalcond2}) for the dilaton $\Phi$. For this,  consider
\begin{eqnarray}
&&I=\int_M \d^DX \sqrt G  e^{-2\Phi} (\hat R-\lambda G+2\hat\nabla \partial \Phi)_{IJ} (\hat R-\lambda G+2\hat\nabla \partial \Phi)^{IJ}
\cr
&&~~
= \int_M \d^DX \sqrt G e^{-2\Phi}  (\hat R_{IJ} \hat R^{IJ}+ 4\hat\nabla_I \partial_J \Phi \hat R^{IJ}+4\hat\nabla_I \partial_J \Phi \hat\nabla^I \partial^J \Phi
\cr
&&\qquad\qquad\qquad -2\lambda \hat R-4\lambda \Delta^2 \Phi+ 4 \lambda^2 D )\,.
\end{eqnarray}
  Clearly if it can be shown that there exists a function $\Phi$
  such that $I=0$, the  conformal invariance condition (\ref{conformalcond2})  will be satisfied. Assuming (\ref{scalecond2}) and after using (\ref{maxprin}) to substitute for the
  $\hat R_{IJ} \hat R^{IJ}$ term, and the Bianchi identity (\ref{b5}) and integration by parts to simplify  the second term,  one finds that
\begin{equation}
I=2\int_M \d^DX \sqrt G \big(\Delta^2 \Phi- (\partial \Phi)^2+\lambda \Phi+{1\over4} \hat R+{1\over24} H^2\big) \big(-\Delta^2 e^{-2\Phi}+D^I(V_I e^{-2\Phi})\big)\,.
\end{equation}
Thus the integrand of $I$ factorises.
Clearly,  $I$ vanishes if either of the two factors  vanishes.

In particular,  $I$ vanishes for $\Phi$ such that
\begin{equation}
\Delta^2 \Phi- (\partial \Phi)^2+\lambda\Phi+{1\over4} \hat R+{1\over24} H^2=0\,.
\end{equation}
In terms of the positive function $W=e^{-\Phi}$, this equation becomes
\be\label{nonlinear}
\left( \Delta^2-{1\over4} \hat R-{1\over24} H^2   +    \lambda \log W \right) W = 0\,. \ee
If $\lambda=0$, the equation becomes linear and asserts that a certain Schr\"{o}dinger operator on the compact manifold $M$ has $W$ as an eigenfunction with zero eigenvalue.
If this is the case, then as $W>0$, $W$ will be the ground state wavefunction of the operator in question.  As a Schr\"{o}dinger operator with a generic potential does not
have a ground state energy of 0, the equation generically does not have a solution for $\lambda=0$.   However, for $\lambda<0$, the left hand side of eqn. (\ref{nonlinear}) is
positive for very small positive $W$ and negative for very large $W$.   That is precisely the situation in which the method of subsolutions and supersolutions can be used
to prove the existence of a solution of an equation of this kind.
For an elementary explanation of this method, see the analysis of eqn. (3.10) in \cite{canonical}.   So for $\lambda<0$, the modified equation (\ref{scalecond2}) of scale invariance does
imply the modified condition (\ref{conformalcond2}) of conformal invariance.    According to Theorem 11.3 in \cite{AD},  the same is true for $\lambda>0$.

\vskip1cm
 \noindent {\it {Acknowledgements}}  We thank C. M. Hull and M Ro\v cek for discussions.
  Research of EW supported in part by NSF Grant PHY-2207584.
 \bibliographystyle{unsrt}

\begin{thebibliography}{99}



\bibitem{friedan}
D. Friedan, ``Nonlinear Models in $2+\epsilon$ Dimensions,'' Phys, Rev. Lett. {\bf 45} (1980) 1057-60.

\bibitem{CFMP}
C. G. Callan, D. Friedan, E. J. Martinec, and M. J. Perry, ``Strings in Background Fields,'' Nucl. Phys. {\bf B262} (1985) 316.

\bibitem{Ts}
E. S.  Fradkin and A. A.  Tseytlin,  ``Effective Field Theory From Quantized Strings,''   Phys. Lett. {\bf  B158} (1985) 316.

\bibitem{Ts2}
E. S. Fradkin and A. A. Tseytlin, ``Quantum String Theory Effective Action,''
 Nucl. Phys. {\bf B261} (1985) 1-27.

\bibitem{CT}
C. G. Callan and  L. Thorlacius, ``Sigma-Models and String Theory,'' in A. Jevicki and C.-I. Tan, eds, {\it Particles, Strings, and Supernovae} (World Scientific, 1989), 795-878.

\bibitem{HullTownsend}
C. M. Hull and P. K. Townsend,
``Finiteness and Conformal Invariance in Nonlinear
$\sigma$  Models,''
 Nucl. Phys. {\bf B274} (1986) 349-362.

\bibitem{Polchinski}
J. Polchinski, ``Scale and Conformal Invariance in Quantum Field Theory,'' Nucl. Phys.  {\bf B303} (1988) 226-236.


\bibitem{Polchinski2}
J.~Polchinski,
``String theory. Vol. 2: Superstring theory and beyond,''
Cambridge University Press, 2007,
ISBN 978-0-511-25228-0, 978-0-521-63304-8, 978-0-521-67228-3


\bibitem{Itsiossfetsossiampos} G.~Itsios, K.~Sfetsos and K.~Siampos, ``Kerr\textendash{}Schild perturbations of coset CFTs as scale invariant integrable \ensuremath{\sigma}-models,'' Nucl. Phys. B \textbf{973}, 115594 (2021) [arXiv:2109.05040 [hep-th]].


\bibitem{Zam}A. Zamolodchikov, `` `Irreversibility' of the Flux of the Renormalization Group in a 2-D Field Theory,'' JETP Lett. {\bf 43} (1986) 430
[Pisma Zh. Eksp. Teor. Fiz. 43, 565 (1986)].


\bibitem{DKST} A. Dymarsky, Z. Komargodsky, A. Schwimmer, and S. Theisen, ``On Scale and Conformal Invariance in Four Dimensions,'' JHEP {\bf 10} (2015) 171,
arXiv:1309.2921.

\bibitem{Bourg}J.-P. Bourguignon, ``Ricci Curvature and Einstein Metrics,''  in D. Ferus and U. Simon, eds., {\it Global Differential Geometry and Global Analysis}, Lecture Notes in Mathematics 838 (Springer, 1981), 42-63.

\bibitem{Perelman} G. Perelman, ``The Entropy Formula For The Ricci Flow and Its Applications,'' math.DG/0211159.

\bibitem{Hamilton}  R. S. Hamilton, ``Three-Manifolds with Positive Ricci Curvature,'' J. Diff. Geom. {\bf 17} (1982) 255.


\bibitem{OSW2} T. Oliynyk, V. Suneeeta, and E. Woolgar, ``A Gradient Flow for Worldsheet Nonlinear Sigma-Models,'' Nucl. Phys. {\bf B739} (2006) 441-458, arXiv:hep-th/0510239.


\bibitem{Huhu} S.~Hu, Z.~Hu and R.~Zhang, ``Generalized Ricci flow and supergravity vacuum solutions,'' Int. J. Mod. Phys. A \textbf{25}, 2535-2549 (2010).


\bibitem{OSW} T. Oliynyk, V. Suneeta, and E. Woolgar, ``Irreversibility of World-sheet Renormalization Group
Flow,''   Phys.Lett. {\bf B610} (2005) 115-121, arXiv:hep-th/0410001.


\bibitem{Tseytlin2}
A. A. Tseytlin, ``On Sigma-model RG Flow, `Central Charge' Action, and Perelman's Entropy,''
arXiv:hep-th/0612296.

\bibitem{Tseytlin3}
A. A. Tseytlin, ``Conditions of Weyl Invariance of the Two-Dimensional Sigma Model From Equations of Stationarity of the `Central Charge' Action,''
Phys. Let. {\bf B194} 63-8.

\bibitem{CP}
G. Curci and G. Paffuti, ``Consistency  Between the String Background Field Equation of Motion and the Vanishing of the Conformal Anomaly,''
Nucl. Phys, {\bf B286} (1987) 399-408.

\bibitem{MorganTian}J. Morgan and G. Tian, {\it Ricci Flow and the Poincar\'{e} Conjecture} (American Mathematical Society).



\bibitem{Hamilton2}
R. S. Hamilton, ``The Ricci Flow on Surfaces,'' Contemporary Mathematics {\bf 71} (1988) 237-61.


\bibitem{Bamler}
R. H. Bamler, ``Recent Developments in Ricci Flow,'' arXIv:2102.12615.

\bibitem{Cao} H.-D. Cao, ``Recent Progress on Ricci Solitons,'' arXiv:0908.2006.









\bibitem{Curtrightzachos}
T.~L.~Curtright and C.~K.~Zachos,
``Geometry, Topology and Supersymmetry in Nonlinear Models,''
Phys. Rev. Lett. \textbf{53}, 1799 (1984)



\bibitem{Howesierra}
P.~S.~Howe and G.~Sierra,
``Two-dimensional supersymmetric nonlinear sigma-models with torsion,''
Phys. Lett. B \textbf{148}, 451-455 (1984)



\bibitem{JGCHMR}
S.~J.~Gates, Jr., C.~M.~Hull and M.~Ro\v cek,
``Twisted Multiplets and New Supersymmetric Nonlinear sigma-models,''
Nucl. Phys. B \textbf{248}, 157-186 (1984)




\bibitem{Braaten}
E.~Braaten, T.~L.~Curtright and C.~K.~Zachos,
``Torsion and Geometrostasis in Nonlinear sigma-models,''
Nucl. Phys. B \textbf{260}, 630 (1985)
[erratum: Nucl. Phys. B \textbf{266}, 748-748 (1986)].

\bibitem{HullHeterotic}
C. M. Hull, ``Compactifications of the Heterotic Superstring,''  Phys. Lett. {\bf B178} (1986) 357.


\bibitem{Hullwitten}
C.~M.~Hull and E.~Witten,
``Supersymmetric Sigma-Models and the Heterotic String,''
Phys. Lett. B \textbf{160}, 398-402 (1985).

\bibitem{StromingerA}
A. Strominger, ``Superstrings WIth Torsion,'' Nucl. Phys. {\bf B274} (1986) 253.

\bibitem{Hullreview}
C.~M.~Hull,
``Lectures on Nonlinear Sigma-Models and Strings,'' in H.-C. Lee et. al., eds,
{\it Superfield Theories}, NATO Sci .Ser. {\bf B160} (1987), pp.
77-168.

\bibitem{Buscher}
T. H. B\"{u}scher, ``Quantum Corrections and Extended Supersymmetry in New $\sigma$ Models,'' Phys. Lett. {\bf B159} (1985) 127-130.

\bibitem{Hull}
C. M. Hull,
``$\sigma$ Model Beta Functions and String Compactifications,''
 Nucl. Phys. {\bf B267} (1986) 266-276.

\bibitem{Hitchin}
N. Hitchin, ``Generalized Calabi-Yau Manifolds,'' Q. J. Math. {]bf 54} (2003) 281-308, arXiv:math/0209099.

\bibitem{Gualtieri} M. Gualtieri, ``Generalized K\"{a}hler Geometry,'' Comm. Math. Phys.
{\bf 331} (2014) 297-331, arXiv:1007.3485.

\bibitem{GFS}
M. Garcia-Fernandez and J. Streets,
{\it Generalized Ricci Flow} (American Mathematical Society, 2021),  arXiv:2008.07004.

\bibitem{GHP}
G. W. Gibbons, S. W. Hawking, and M. J. Perry,  ``Path Integrals and the Indefiniteness of
the Gravitational Action,''  Nucl. Phys. {\bf B138} (1978) 141-50.

\bibitem{LeeParker}
J. M. Lee and T. H. Parker, ``The Yamabe Problem,'' Bull. Amer. Math. Soc. {\bf 17} (1987)  37-91.

\bibitem{AA}
A. Ahmadain and A. Wall, ``Off-Shell Strings I: $S$-Matrix and Action,''
arXiv:2211.08607.



\bibitem{Zumino}
B.~Zumino,
``Supersymmetry and K\"{a}hler Manifolds,''
Phys. Lett. B \textbf{87}, 203 (1979).



\bibitem{LAGDF}
L.~Alvarez-Gaume and D.~Z.~Freedman,
``Geometrical Structure and Ultraviolet Finiteness in the Supersymmetric sigma-model,''
Commun. Math. Phys. \textbf{80}, 443 (1981).


\bibitem{Curtrightfreedman} T.~L.~Curtright and D.~Z.~Freedman, ``Nonlinear $\sigma$ Models With Extended Supersymmetry in Four-dimensions,'' Phys. Lett. B \textbf{90}, 71 (1980) [erratum: Phys. Lett. B \textbf{91}, 487 (1980)].







\bibitem{HullLocal}
C. M. Hull, U. Lindstrom, M. Ro\v cek, and R. v. Unge, ``Generalized Calabi-Yau Metric and Generalized Monge-Ampere Equation,''
  JHEP {\bf 08} (2010), 060, arXiv:1005.5658.

\bibitem{ST1}
J.~Streets and G.~Tian,
``Regularity results  for pluriclosed flow,''
[arXiv:1008.2794 [math.DG]].


\bibitem{ST2}
J.~Streets and G.~Tian,
``Generalized Kaehler geometry and the pluriclosed flow,''
Nucl. Phys. B \textbf{858}, 366-376 (2012)
[arXiv:1109.0503 [math.DG]].


\bibitem{HoweGP}
P.~S.~Howe and G.~Papadopoulos,
``Twistor spaces for HKT manifolds,''
Phys. Lett. B \textbf{379}, 80-86 (1996).


\bibitem{IvanovGP}
S.~Ivanov and G.~Papadopoulos,
``Vanishing Theorems and String Backgrounds,''
Class. Quant. Grav. \textbf{18}, 1089-1110 (2001)
[arXiv:math/0010038 [math.DG]].


\bibitem{Spindel}
P.~Spindel, A.~Sevrin, W.~Troost and A.~Van Proeyen,
``Extended Supersymmetric sigma-models on Group Manifolds. 1. The Complex Structures,''
Nucl. Phys. B \textbf{308}, 662-698 (1988).

\bibitem{OP}
A. Opferman and G. Papadopoulos, ``Homogeneous HKT and QKT manifolds,''
arXiv:math-ph/9807026.

\bibitem{Verbitsky}
M. Verbitsky, ``Hyper-K\"{a}hler Manifolds With Torsion Obtained from Hyperholomorphic Bundles,''
Math. Research Lett. {\bf 10} (2003) 501-13, arXiv:math/0303129.













\bibitem{HoweGP1}
P.~S.~Howe and G.~Papadopoulos,
``Ultraviolet Behavior of Two-dimensional Supersymmetric Nonlinear $\sigma$ Models,''
Nucl. Phys. {\bf B289}, 264-276 (1987).

\bibitem{Sevrin}
A. Sevrin, W. Troost, and A. van Proeyen, ``Superconformal Algebras in Two Dimensions With $\N=4$,'' Phys. Lett {\bf B208} (1988) 447-50.


\bibitem{HoweGP2}
P.~S.~Howe and G.~Papadopoulos,
``Further Remarks on the Geometry of Two-dimensional Nonlinear $\sigma$ Models,''
Class. Quant. Grav. \textbf{5}, 1647-1661 (1988).


\bibitem{RocekSS}
M.~Rocek, K. Schoutens, and A. Sevrin, ``Off-Shell WZW Models in Extended Superspace,''
Phys, Lett. {\bf B265} (1991) 303-6.

\bibitem{IKR}
I. T. Ivanov, B.-b. Kim, and M. Ro\v cek,
``Complex Structures, Duality and WZW Models in Extended Superspace,'' Phys. Lett. {\bf B343} (1995) 133-143,
: hep-th/9406063.


\bibitem{ADRS}
J.P. Ang, S. Driezen, M.  Ro\v cek, A. Sevrin, ``Generalized K\"{a}hler Structures on Group Manifolds and $T$-Duality,'' JHEP {\bf 05} (2018) 189, arXiv:1804.03259.



 \bibitem{Ivanov} E.~Ivanov and A.~Sutulin, ``Sigma models in (4,4) harmonic superspace,'' Nucl. Phys. B \textbf{432}, 246-280 (1994) [erratum: Nucl. Phys. B \textbf{483}, 531-531 (1997)]
[arXiv:hep-th/9404098 [hep-th]].

\bibitem{GP}
G.~Papadopoulos,
``Elliptic monopoles and (4,0) Supersymmetric Sigma-models With Torsion,''
Phys. Lett.  \textbf{B356} (1995), 249-255
[arXiv:hep-th/9505119.


\bibitem{Tod}
T.~Chave, G.~Valent and K.~P.~Tod,
``(4,0) and (4,4) sigma-models with a triholomorphic Killing vector,''
Phys. Lett. B \textbf{383} (1996), 262-270.



\bibitem{BaSw}
B. Banos and A. Swann, ``Potentials for Hyper-K\"{a}hler Metrics With Torsion,'' arXiv:math/0402366.


\bibitem{Strominger}
A.~Strominger,
``Superstrings with Torsion,''
Nucl. Phys. B \textbf{274} (1986) 253.



\bibitem{Duff}
M.~J.~Duff and J.~X.~Lu,
``Elementary Five-Brane Solutions of D = 10 Supergravity,''
Nucl. Phys. B \textbf{354} (1991) 141-153.

\bibitem{CHS}C. C. Callan,  J.  A. Harvey,  and A. E. Strominger,  ``Supersymmetric String Solitons,'' in J. A. Harvey et. al., eds,  {\it String Theory and Quantum Gravity `91} (World Scientific, 1992),
arXiv:hep-th/9112030.

\bibitem{canonical}
E. Witten, ``A Note on the Canonical Formalism for Gravity,'' arXiv:2212.08270.

\bibitem{AD}
A. Derdzinski, ``Ricci Solitons,''  arXiv:1712.06055.

\end{thebibliography}

\end{document}